\documentclass[twocolumn]{aastex631}

\usepackage{amssymb}

\newcommand{\kzz}{$K_{zz}$}
\newcommand{\fsed}{$f_{\rm sed}$}
\newcommand{\teff}{$T_{\rm eff}$}

\begin{document}

\title{The Sonora Substellar Atmosphere Models. III. Diamondback: Atmospheric Properties, Spectra, and Evolution for Warm Cloudy Substellar Objects}

\author[0000-0002-4404-0456]{Caroline V. Morley}
\affiliation{Department of Astronomy, University of Texas at Austin, 2515 Speedway, Austin, TX 78712, USA}

\author[0000-0003-1622-1302]{Sagnick Mukherjee}
\affiliation{Department of Astronomy \& Astrophysics, University of California Santa Cruz, Santa Cruz, CA, USA}
\author[0000-0002-5251-2943]{Mark S. Marley}
\affiliation{Department of Lunar and Planetary Sciences, University of Arizona, Tucson, Arizona, USA}

\author[0000-0002-9843-4354]{Jonathan J. Fortney}
\affiliation{Department of Astronomy \& Astrophysics, University of California Santa Cruz, Santa Cruz, CA, USA}

\author[0000-0001-6627-6067]{Channon Visscher}
\affiliation{Chemistry \& Planetary Sciences, Dordt University, Sioux Center IA 51250}\affiliation{Center for Extrasolar Planetary Systems, Space Science Institute, Boulder, CO 80301}

\author[0000-0003-3444-5908]{Roxana Lupu}
\affiliation{Eureka Scientific, Inc., Oakland, CA 94602}

\author[0000-0002-4088-7262]{Ehsan Gharib-Nezhad}
\affiliation{Space Science and Astrobiology Division, NASA Ames Research Center, Moffett Field, CA, 94035 USA}

\author[0000-0002-5113-8558]{Daniel Thorngren}
\affiliation{Department of Physics \& Astronomy, Johns Hopkins University, Baltimore, MD, USA}

\author[0000-0001-9333-4306]{Richard Freedman}
\affiliation{SETI Institute, Mountain View, CA, USA}

\author[0000-0003-1240-6844]{Natasha Batalha}
\affiliation{Space Science and Astrobiology Division, NASA Ames Research Center, Moffett Field, CA, 94035 USA}

\begin{abstract}
We present a new grid of cloudy atmosphere and evolution models for substellar objects. These models include the effect of refractory cloud species, including silicate clouds, on the spectra and evolution. We include  effective temperatures from 900 to 2400 K and surface gravities from log g=3.5--5.5, appropriate for a broad range of objects with masses between 1 and 84 M$_J$. Model pressure--temperature structures are calculated assuming radiative--convective and chemical equilibrium. We consider the effect of both clouds and metallicity on the atmospheric structure, resulting spectra, and thermal evolution of substellar worlds. We parameterize clouds using the \citet{AM01} cloud model, including cloud parameter \fsed\ values from 1--8; we include three metallicities ($-0.5$, 0.0, and $+0.5$). Refractory clouds and metallicity both alter the evolution of substellar objects, changing the inferred temperature at a given age by up to 100--200 K. We compare to the observed photometry of brown dwarfs, finding broad agreement with the measured photometry. We publish the spectra, evolution, and other data products online with open access. 

\end{abstract}

\section{Introduction}

Brown dwarfs and giant exoplanets evolve over their lifetimes, continuously changing as they cool. They spend the first part of their lives as `hot' objects with largely cloud-free atmospheres, their spectra shaped by metal hydrides and oxides, neutral metals, water vapor, and carbon monoxide \citep{Kirkpatrick05, Lodders06}. As they cool, the most refractory species condense, forming dusty clouds likely composed of aluminum-bearing oxides (e.g., Al$_2$O$_3$), metals (e.g., Fe), and silicates (e.g., MgSiO$_3$, Mg$_2$SiO$_4$) \citep{Lunine86, Burrows97, Chabrier00b, Marley02}. 

These clouds clear rapidly at the L/T transition, at effective temperatures (\teff) around 1300 K, around the same temperature as CH$_4$ begins to be observed in near-infrared spectra \citep{Burgasser02a}. While atmospheric models naturally predict that atmospheres become more clear when the silicate clouds sink below the photosphere \citep{Saumon08}, the rapid speed of this transition is still not well-understood or reproduced by models. T dwarfs remain largely clear of clouds, though trace species can potentially create low-optical-depth clouds in later T and early Y dwarfs \citep{Morley12}. The coolest Y dwarfs, cooler than 375 K, likely condense water into clouds that become thicker as the object cools further \citep{Burrows03b, Morley14a}. 

\subsection{Observational evidence for clouds in L dwarfs}

Classically, the observational evidence for cloud opacity in L dwarfs comes from their red near-infrared spectra. Models that include a layer of largely silicate-based clouds match near-infrared spectra significantly better than cloud-free models \citep{Cushing06, Burrows06, Cushing08, Stephens09}. A growing line of evidence suggests that lower-gravity objects including directly-imaged planets have thicker clouds that make their near-infrared spectra and colors even redder \citep[e.g.,][]{Allers13, Faherty16, Marois08, Currie11}. Further, it appears that clouds in directly-imaged planets and the lowest mass brown dwarfs may persist to lower effective temperatures (i.e., the dramatic clearing of clouds at the L/T transition is actually gravity-dependent) \citep{Marley10, Currie11}. 

The silicate feature at 9--10 \micron\ provides more direct evidence that silicates are forming dusty clouds in the upper atmosphere. This feature was first seen in Spitzer/IRS spectra \citep{Cushing06}. \citet{Luna21} found using forward models that these spectra were most consistent with amorphous enstatite (MgSiO$_3$) grains in the upper atmosphere (1--10 mbar). \citet{Burningham21} found using retrievals that considered over 60 different cloud combinations that the clouds in L dwarf 2M2224-0158 were most consistent with enstatite and quartz. In contrast, studying the younger, lower-mass SIMP0136, \citet{Vos23} found that the cloud composition is more consistent with forsterite (Mg$_2$SiO$_4$) clouds. 

Recently, the full Spitzer/IRS sample has been re-reduced and published by \citet{Suarez22}. Their analysis finds that silicate absorption is first seen at spectral type L2, is strongest in L4-L6 dwarfs, and disappears beyond L8. Furthermore, they find that the shape of the silicate feature is gravity-dependent, appearing consistent with amorphous enstatite for older objects and with iron-bearing pyroxene (Mg$_x$Fe$_{1-x}$SiO$_3$) for younger objects.  \

\subsection{A need for new cloudy atmosphere and evolution models }

Since the publication of \citet{Saumon08}, there have been a number of steps forward in our understanding of substellar atmospheres. Importantly for generating spectra of substellar atmospheres, many molecules have more complete and accurate line lists, notably CH$_4$ \citep{Yurchenko14, Hargreaves20}, NH$_3$ \citep{Yurchenko11, Wilzewski16}, metal oxides \citep{McKemmish16, McKemmish2019TiO, GharibNezhad2021}, metal hydrides \citep{GharibNezhad2013MgH, Hargreaves2010FeH}, and the alkali metals \citep{Allard2019}. In recent years water and methane line lists for high-temperature atmospheres have been substantially updated using ab initio and laboratory-measured spectra. TiO line positions and spectral intensities have been modified. Temperature-dependent pressure-broadening coefficients for brown dwarf atmospheres dominated by H$_2$/He gases have been calculated, and several bands of major and minor isotopologues have been added to the recent line lists for ExoMol and HITRAN/HITEMP, which are included in this grid model. Please see \citep{hitran2020, ExoMol2020, GharibNezhad2021} for further details on the line lists and their computation procedures.

Since 2008, we have also learned that the cloud-clearing of the L/T transition is likely gravity-dependent, and evolution models should be updated to accurately reflect this. Furthermore, with the discovery of directly-imaged planets since 2008, a need for higher metallicity ($\sim3\times$ solar) models, akin to Jupiter's metallicity, are needed. 

Our team has published cloud-free models that include the opacity updates above for a broad range of metallicities and C/O ratios. Models with chemical equilibrium were published first as `Sonora Bobcat' \citep{Marley21}. Next, \citet{Karalidi21} included chemical disequilibrium in these models self-consistently, publishing the `Sonora Cholla' grid. And most recently, the chemical disequilibrium grid was expanded in `Sonora Elf-Owl' to include a broad range of metallicities and C/O ratios (Mukherjee et al. submitted). 

Other teams have also published atmosphere models that include many of these upgrades.
The BT-Settl models are widely used for late M, L, and T dwarfs and include a parameterized cloud model \citep{Allard14}. 
The Exo-REM models focus on the L and T dwarfs \citep{Charnay18}. They include a new `simple microphysics' cloud model and also disequilibrium chemistry, which becomes important for late L and early T dwarfs. \citet{Lacy23} covers similar physics---clouds plus disequilibrium chemistry---for colder objects than we consider here with water ice clouds. The ATMO 2020 models \citep{Phillips20} are comparable to our `Bobcat' models, with fully cloud-free atmospheres. Last, \citet{Tremblin15,Tremblin19} have considered models that lack clouds but tweak the pressure--temperature profiles at the relevant layers to compensate, invoking diabatic processes that change the temperature structure. 

The atmosphere models described above are typically not coupled self-consistently to the thermal evolution, but the thermal evolution can be sensitive to the atmospheric boundary condition used. Because the overall photospheric opacity is higher in the presence of clouds, a cloudy radiative--convective atmosphere with a given \teff\ is hotter at every pressure level than an equivalent cloudless atmosphere. This means that a cloudy atmosphere’s deep adiabat is also hotter and the interior is higher entropy at every \teff. The evolution through time, as an object cools to progressively cooler adiabats, is therefore altered by the presence of clouds. Realistic evolution models thus require a treatment of clouds to compute the correct boundary condition linking \teff, gravity, metallicity, and the deep adiabat.

The cloudless \citet{Phillips20} models have been used as boundary conditions for new evolution calculations \citep{Chabrier23} with an updated equation of state. 

Here, we publish the first of the cloudy Sonora grids, appropriate for warmer (L and early T) brown dwarfs and directly-imaged planets. These are unique in the current evolution modeling landscape in the fact that they use cloudy atmosphere models for the boundary conditions at  L dwarf temperatures. 
We dub our new models `Sonora Diamondback' after the Western Diamondback rattlesnake, \emph{Crotalus atrox}, common in the Sonoran desert and Central Texas. 

\begin{deluxetable}{ccc}
\tabletypesize{\footnotesize}
\tablecolumns{3}
\tablewidth{10pt}
\tablecaption{ Model Grid Parameters \label{table:modelgrid}}
\tablehead{
\colhead{Parameter} & \colhead{Range } & \colhead{Step} }
\startdata
$T_{\rm eff}$ (K) & 900--2400  & 100 \\ 
log g (cgs) & 3.5--5.5  & 0.5 \\
\ [M/H] (dex) &  $-$0.5 to +0.5  & 0.5  \\
$f_{\rm sed}$ &  [1, 2, 3, 4, 8, nc] &  - \\
\enddata
\vspace{-0.8cm}
\end{deluxetable}

\begin{table*}
    \centering
    \begin{tabular}{c|c}
         C$_2$H$_2$ & \citet{hitran2012}\\
         C$_2$H$_4$ & \citet{hitran2012}\\
         C$_2$H$_6$ & \citet{hitran2012} \\
         CH$_4$ & \citet{Yurchenko:2013}, \citet{Yurchenko:2014}, \citet{STDS}, \citet{Pine:1992}, \\
            & \citet{Hargreaves20}, 
         \citet{GharibNezhad2021}\\
         CO &  \citet{HITEMP2010,HITRAN2016,li15rovibrational}\\
         CO$_2$ &  \citet{HUANG2014reliable}\\
         CrH &  \citet{Burrows02_CrH}, computed in \citet{GharibNezhad2021}\\
         Fe &  \citet{Ryabchikova2015,oBrian1991Fe,Fuhr1988Fe, Bard1991Fe,Bard1994Fe} \\
         FeH &  \citet{Dulick2003FeH, Hargreaves2010FeH} \footnote{We divide the FeH opacities in H-band (at 1.6 $\mu$m) by 3 to better reproduce observed spectra.} \\
         H$_2$ & \citet{HITRAN2016} \\
         H$_3^+$ &  \citet{Mizus2017H3p}\\
         H$_2$--H$_2$ & \citet{Saumon12} with added overtone from \citet{Lenzuni1991h2h2} Table 8\\
         H$_2$--He &  \citet{Saumon12} \\
         H$_2$--N$_2$ &  \citet{Saumon12} \\
         H$_2$--CH$_4$ &  \citet{Saumon12} \\
         H$_2^-$  &  \citet{bell1980free}\\
         H$^-$ bf &  \citet{John1988H}\\
         H$^-$ ff &  \citet{Bell1987Hff}\\
         H$_2$O &  \citet{Polyansky2018H2O}\\
         H$_2$S &  \citet{azzam16exomol}\\
         HCN &  \citet{Harris2006hcn,Barber2014HCN,hitran2020}\\
         LiCl &  \citet{Bittner2018Lis}\\
         LiF &  \citet{Bittner2018Lis}\\
         LiH & \citet{Coppola2011LiH} \\
         MgH & \citet{Yadin2012MgH,GharibNezhad2013MgH} computed in \citet{GharibNezhad2021}\\
         N$_2$ &  \citet{hitran2012}\\
         NH$_3$ &  \citet{yurchenko11vibrationally,Wilzewski16} \\
         OCS &  \citet{HITRAN2016}\\
         PH$_3$ & \citet{sousa14exomol} \\
         SiO &  \citet{Barton2013SiO, GharibNezhad2021} \\
         TiO & \citet{McKemmish2019TiO} computed in \citet{GharibNezhad2021}\\
         VO &   \citet{McKemmish16} computed in \citet{GharibNezhad2021}\\
         Li,Na,K &  \citet{Ryabchikova2015,Allard2007AA, Allard2007EPJD,Allard2016, Allard2019}, as compiled in \citet{Molliere19b} \\
         
    \end{tabular}
    \caption{References of gaseous opacities used for calculating the atmospheric models and resulting spectra in this work.}
    \label{tab:opa_tab}
\end{table*}

\subsection{This work}

We present a new set of 1440 atmospheric spectra designed to be appropriate for brown dwarfs and directly-imaged planets, including three different atmospheric metallicities. We use these atmospheric models as the boundary conditions for models of the evolution of substellar objects including the emergence and disappearance of clouds in their spectra. We show how these models differ from prior generations of models. 

\section{Methods}

We generate new one-dimensional atmosphere and evolution models appropriate for brown dwarfs and directly-imaged planets. Our grid is summarized in Table \ref{table:modelgrid}. It extends from 900--2400 K, with log g from 3.5 to 5.5, metallicities ([M/H]) of $-$0.5, 0.0, and $+$0.5 (scaling relative to the Sun, \citet{Lodders09}). All models assume solar C/O  ratio (=0.458 \citet{Lodders09}). We include a range of cloud thickness, as parameterized by \fsed\ (see Section \ref{sec:clouds}).  

\subsection{Atmosphere model}

We calculate one-dimensional pressure--temperature (P--T) profiles which are in radiative--convective and chemical equilibrium. The atmosphere model we use is described more extensively in \citet{Mckay89, Marley96, Burrows97, Marley99, Marley02, Saumon08}, and has been used to model a variety of brown dwarf and exoplanet atmospheres \citep{Fortney08b, Marley10, Morley12, Morley14a, Morley14b, Marley21, Karalidi21}. 

The thermal radiative transfer is determined using the ``source function technique'' \citep{Toon89}. The gas opacity is calculated using correlated-k coefficients to increase calculation speed; our opacity database has been described in more extensive detail in \citet{Freedman08, Freedman14,Mukherjee2023}, and the sources of the line lists are summarized in Table \ref{tab:opa_tab}.



\subsection{Clouds} \label{sec:clouds}

To include clouds, in a simple but physically plausible way, we follow the mass-balance approach of \citet{AM01}. We assume a wide log-normal particle size distribution with geometric size distribution $\sigma_g$=2.  

\subsubsection{Cloud Optical Properties}

The cloud opacity is included as Mie scattering of spherical cloud particles in each atmospheric layer. We choose amorphous silicates (rather than crystalline) based on the empirically determined compositions from \citet{Luna21, Burningham21}: MgSiO$_3$: \citet{Dorschner95};  Mg$_2$SiO$_4$: \citet{Jager03}, Fe: \citet{Kitzmann18, Palik91},  Al$_2$O$_3$: \citet{Koike95}. 

\begin{figure*}
    \centering
    \includegraphics[width=7in]{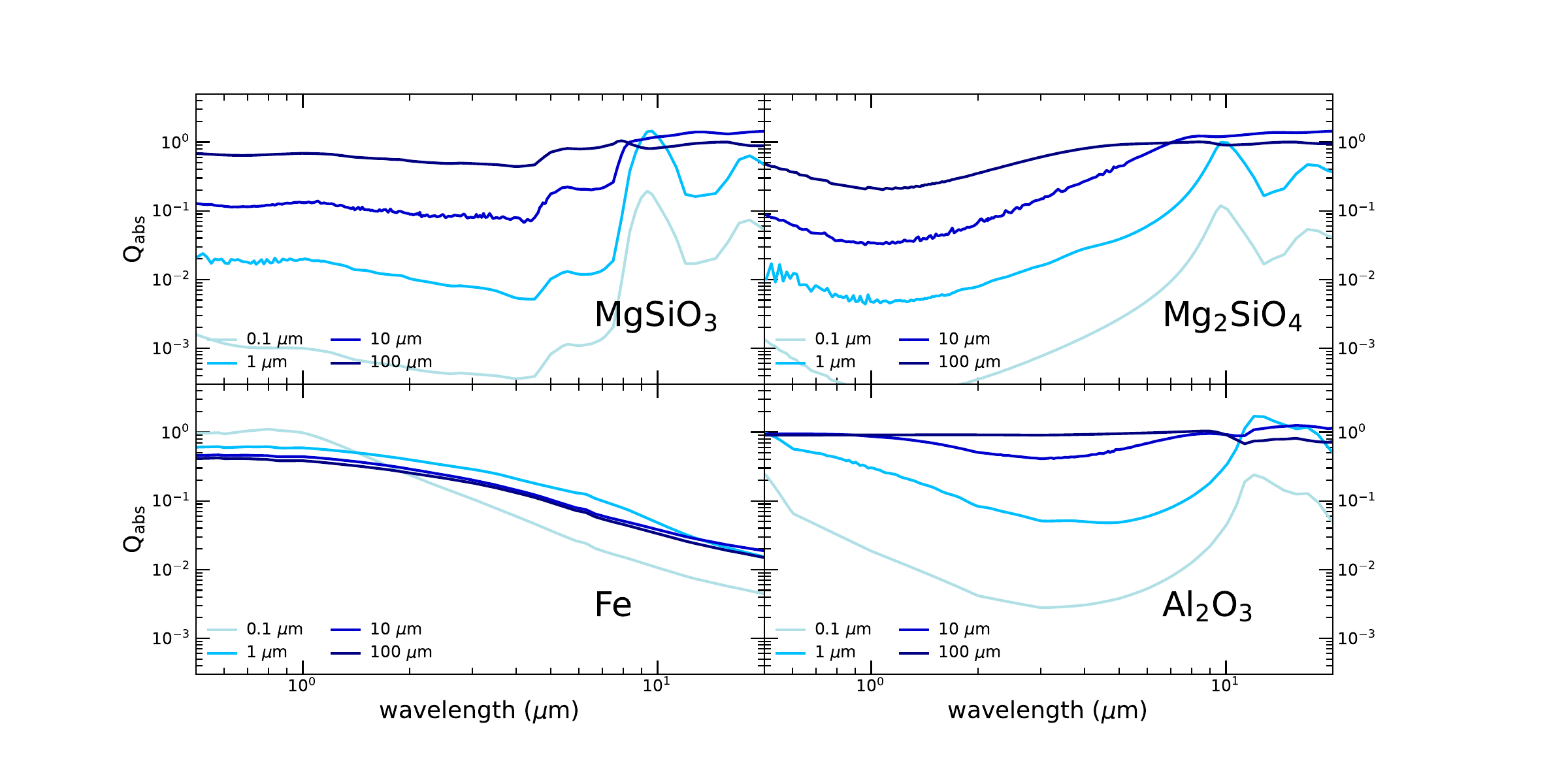}
    \caption{Mie scattering absorption efficiencies, Q$_{\rm abs}$, of the four clouds included in these models: MgSiO$_3$, Mg$_2$SiO$_4$, Fe, and Al$_2$O$_3$. Note the large features for silicates at $\sim$10 $\mu$m and for Al$_2$O$_3$ at $\sim$13 $\mu$m. For each cloud species, Q$_{\rm abs}$ has larger amplitude features for smaller particles.  }
    \label{fig:miescat}
\end{figure*}

We show examples of the cloud absorption properties in Figure \ref{fig:miescat}, calculating the Mie scattering absorption $Q_{\rm abs}$ at four representative particle sizes (0.1, 1, 10, and 100 \micron). As in prior works, we see that silicates with 0.1-1 \micron\ particles have strongly wavelength-dependent absorption, with a peak at $\sim$10 \micron. Iron typically has few distinctive absorption properties. Al$_2$O$_3$ with 0.1-1 \micron\ particles has an absorption feature at 11-13 \micron. 

\subsubsection{Cloud Saturation Vapor Pressures}

We calculate the condensation of cloud material using the saturation vapor pressure of the limiting gas for each species. We use metallicity-dependent saturation vapor pressure formulae derived from the chemistry models between metallicities of -1.0 and +1.0 dex, and are most accurate within those ranges. In this approach, the condensation condition is met when the atmospheric partial pressure of a limiting gas species $j$ ($p_j$) becomes greater than the  saturation vapor pressure of $j$ ($p_j'$)
\begin{equation}\label{equation: condensation condition}
p_{j} \geq p_j',
\end{equation} 
where $p_j$ is the atmospheric partial pressure of $j$ and $p_j'$ is the temperature-dependent vapor pressure of  $j$ above the condensate. Above the cloud, the partial pressure of $j$ is given by $p_j'$.

In a change from previous models \citep[e.g.,][]{Saumon08}, but in accord with prior chemistry modeling \citep{Visscher10}, we assume that both forsterite and enstatite can condense. Forsterite (Mg$_2$SiO$_4$) is assumed to condense at deeper pressures than enstatite (MgSiO$_3$). We calculate the saturation vapor pressure of Mg$_2$SiO$_4$ assuming that Mg is the limiting element and monatomic Mg is the dominant Mg-bearing species: 
\begin{equation}
    \log_{10}(p_{\rm Mg}') \approx \frac{-32488}{T} + 14.88 - 0.2\log_{10}P_{\rm t} - 1.4 [\rm{M}/\rm{H}] 
\end{equation}
where the vapor pressure $p_{\rm Mg}'$ is in bar, $T$ is the temperature of the atmospheric layer in K, $P_{\rm t}$ is the total gas pressure in bar, and $[\rm{M}/\rm{H}]$ is the metallicity in log units (e.g., solar is 0.0). 

We assume that the MgSiO$_3$ cloud forms at lower pressures, above the Mg$_2$SiO$_4$ cloud. Since the forsterite cloud has removed both Mg and Si from the vapor phase, we find that Si is now the limiting element for enstatite cloud formation and that SiO is the dominant Si-bearing species at temperatures near the cloud base. For MgSiO$_3$, the saturation vapor pressure is thus approximated by: 
\begin{equation}
        \log_{10}p_{\rm SiO}' \approx \frac{-28665}{T} + 13.43 
\end{equation}
Monatomic Fe vapor is the dominant Fe-bearing species and condenses directly from the gas phase (instead of via a chemical reaction). Its  vapor pressure is given by 
\begin{equation}
        \log_{10}p_{\rm Fe}' \approx \frac{-20995}{T} + 7.09 
\end{equation}

Finally, we assume that Al$_2$O$_3$ is the dominant aluminum-bearing cloud. The actual condensation sequence of the most refractory materials may be more complex \citep[see, e.g.,][]{ Wakeford17}, but this approach captures the bulk condensation of dusty, refractory species. The saturation vapor pressure for the Al$_2$O$_3$ cloud can be approximated by considering monatomic Al as the limiting Al-bearing species:
\begin{equation}
        \log_{10}p_{\rm Al}' \approx \frac{-41481}{T} + 15.24 - 1.5 [\rm{M}/\rm{H}]
\end{equation}

`Condensation curves' can be calculated for different metallicities using these saturation vapor pressure relationships and finding the layers of the atmosphere where the abundance of the limiting condensible species is equal to the saturation vapor pressure (i.e., $p_j \geq p_j'$).

\begin{figure}
    \centering
    \includegraphics[width=3.5in]{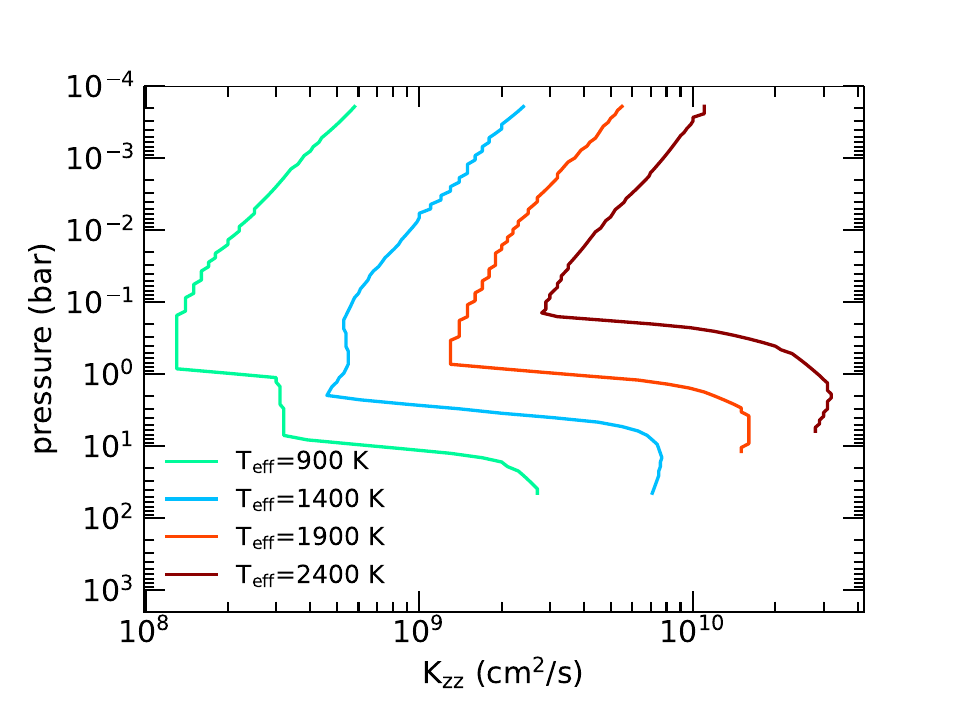}
    \caption{Representative eddy diffusion coefficients, \kzz, which parameterizes vertical mixing of the cloud material. Profiles are taken for models with effective temperatures spanning our grid, and  have log g=4.5, \fsed=3, and [M/H]=0.0. }
    \label{fig:kzz}
\end{figure}

\subsubsection{Eddy Mixing Parameterization}

The cloud model, based on the method presented in \citet{AM01}, balances the upward turbulent mixing of condensate and vapor with the downward transport of condensed material due to sedimentation in the atmosphere, via equation (4) from that work: 

\begin{equation}
    -K_{zz} \frac{\partial q_t}{\partial z} - f_{\rm sed} w_* q_c =0
\end{equation}
where $K_{zz}$ is the eddy diffusion coefficient, $w_*$ is the convective velocity, $q_t$ is the mixing ratio of vapor plus condensate, $q_c$ is the mixing ratio of condensate, $z$ is the height in the atmosphere, and $f_{\rm sed}$ is the sedimentation efficiency and the key `free' parameter in the model.  

$K_{zz}$ is calculated using mixing length theory following \citet{Gierasch85}. A key parameter in that model is the convective heat flux in each layer of the atmosphere, $F_{\rm conv}=\sigma T_{\rm eff}^4$ in a fully convective atmosphere. We modify the calculation of the convective heat flux, subtracting the energy transported by radiation in each layer of the atmosphere, $F_{\rm conv}'=\sigma T_{\rm eff}^4 - F_{\rm rad}$. We approximate the effect of convective overshoot by not allowing the convective heat flux to decrease in a layer $z$ by more than $\frac{1}{3} (P_z / P_{z+1}$). The mixing length is set equal to the scale height in the atmosphere. In a modification from prior models \citep{Saumon08, Morley12}, we assert that the convective heat flux should not increase with decreasing pressure, which could occur numerically in detached convective zones in prior models. 

Example $K_{zz}$ profiles are shown in Figure \ref{fig:kzz}. Typically, hotter models have stronger mixing (larger \kzz). The mixing is strong in the deepest layers of the atmosphere and decreases within the convective zone. It reaches a local minimum at the radiative-convective boundary and then increases again in the upper atmosphere. This pattern is broadly similar to more detailed 3D models \citep[e.g.,][]{Parmentier13b}.

\subsection{Evolution models}\label{sec:evolution}

We calculate the thermal evolution of substellar objects using an extension of the thermal evolution model described in \citet{Thorngren16}. This model solves the typical set of structure and evolutionary equations, as in \citet{Saumon08}. These include the mass conservation,
\begin{equation}\label{eq:mass_conserv}
    \frac{d{r}}{d{m}}= \frac{1}{4{\pi}r^2\rho}
\end{equation}
where $m$ is the mass enclosed within the radial distance $r$ and $\rho$ is the density. We also assume that the object is always in hydrostatic equilibrium described by,
\begin{equation}\label{eq:hydrostatic}
    \frac{dP}{dm}= - \frac{Gm}{4{\pi}r^4}
\end{equation}
where $P$ is the pressure. The luminosity of the object is calculated using,
\begin{equation}\label{eq:luminosity}
    \left(L-\int_{0}^{M}{\epsilon_{nuc}dm}\right)dt = -\int_{0}^{M}{TdSdm}
\end{equation}
where $L$ is the luminosity of the object and $M$ is its total mass. $\epsilon_{nuc}$ is the rate of nuclear burning per unit mass, $S$ is the specific entropy, and $T$ is the temperature. We assume that the interior of the object is fully convective, well mixed, and thus isentropic at any given time. The source of the nuclear burning term $\epsilon_{nuc}$ are the $p-p$ chain reactions and deuterium burning, as outlined in \citet{Saumon08}.

We use the H-He mixture equation of state (EOS) presented in \citet{Chabrier19} for all our evolutionary calculations. To reflect the atmospheric metallicity of the object in its interior as well, we mix the 50/50 rock--ice mixture equation of state from \citet{ANEOS} with the H-He equation of state, via the additive volume rule, by assuming an uniform distribution of the same amount of metal in the interior as the atmosphere. The evolutionary calculation can be divided between three components -- the interior structure calculation, nuclear burning calculation, and the time evolution calculation.  A converged interior structure model uses the EOS and the depth-dependent adiabatic temperature gradient to determine the pressure, density, and temperature structure throughout an interior model over 1000 mass shells.

The temperature structure of the interior at a certain time $t$ is used to calculate the nuclear luminosity term $\epsilon_{nuc}$ using the H, $^2$H, and He mass fractions. All models are initiated with a H, $^2$H, and He mass fractions of 0.7112, 2.88$\times10^{-5}$, and 0.2888, respectively. We use the same treatment of nuclear reaction rates as described in \citet{Saumon08} including screening factors from ionic and electronic interactions. Based on the rate of nuclear burning, the amount of H, $^2$H, and He are also changed in each time step. If nuclear burning takes place in the interior, it leads to depletion of H and $^2$H over time in the interior and increase in the amount of He.

The time evolution calculation is based on Equation \ref{eq:luminosity} as it gives us the time interval $dt$ for the entropy of the interior to change by an amount $dS$. Each object is initiated with a very high entropy at $t=0$. In each time step, the entropy of the interior is used to determine the interior structure of the object using the calculation described above. The converged temperature structure along with the equation of state is used to determine the $T_{10}$ from the interior model, which is defined as the temperature at the pressure level of 10 bars. The pressure-temperature profiles from the atmospheric grid are also used to determine $T_{10}$ as a function of {\teff}, $\log (g)$, {\fsed}, and  [M/H]. The $T_{10}$ derived from the atmospheric model grid acts as a boundary condition for the interior model. For the $\log (g)$ and $T_{10}$ values calculated from the interior model, we search for the atmospheric {\teff} which produces the same $T_{10}$ at the $\log (g)$ of the time step. 
The luminosity of the object at that time step is then calculated as $L=4{\pi}R^2\sigma$\teff$^4$. 
The calculated $L$, the nuclear luminosity term $\epsilon_{nuc}$, and the temperature structure of the interior are used as inputs in Equation \ref{eq:luminosity} to determine the $dt$ for a certain amount of decrease in the entropy $dS$. This time evolution step is then again repeated for the next time step $t+dt$.

We verify the accuracy of our new models by comparing to the \citet{Saumon08} models and Sonora Bobcat models \citep{Marley21}, finding excellent agreement with prior approaches. 

We calculate the evolution for each metallicity (0.0, $+$0.5, $-$0.5) for three scenarios: 
\begin{enumerate}
    \item \emph{Cloud-free}: These models use cloud-free models as boundary conditions. We use the new cloud-free models presented here for 900--2400 K and the cloud-free Sonora Bobcat models from \citet{Marley21} from 200--850 K. 
    \item \emph{Hybrid}: These models use cloudy models as boundary conditions above 1300 K and cloud-free models below 1300 K. We use the new cloud-free and cloudy models presented here for 900--2400 K and the cloud-free Sonora Bobcat models from \citet{Marley21} from 200--850 K. 
    \item \emph{Hybrid-grav}: These models include a gravity dependent cloud-clearing at the L/T transition. Specifically, for log g=3.5 we transition from cloudy to cloud-free at 1000 K; for log g=4.0 at 1100 K; for log g=4.5 at 1200 K; for log g=5.0 at 1300 K; and for log g=5.5 at 1400 K. We use the new cloud-free and cloudy models presented here for 900--2400 K and the cloud-free Sonora Bobcat models from \citet{Marley21} from 200--850 K. 
    
\end{enumerate}




\section{Results}

\begin{figure}
    \includegraphics[width=3.5in]{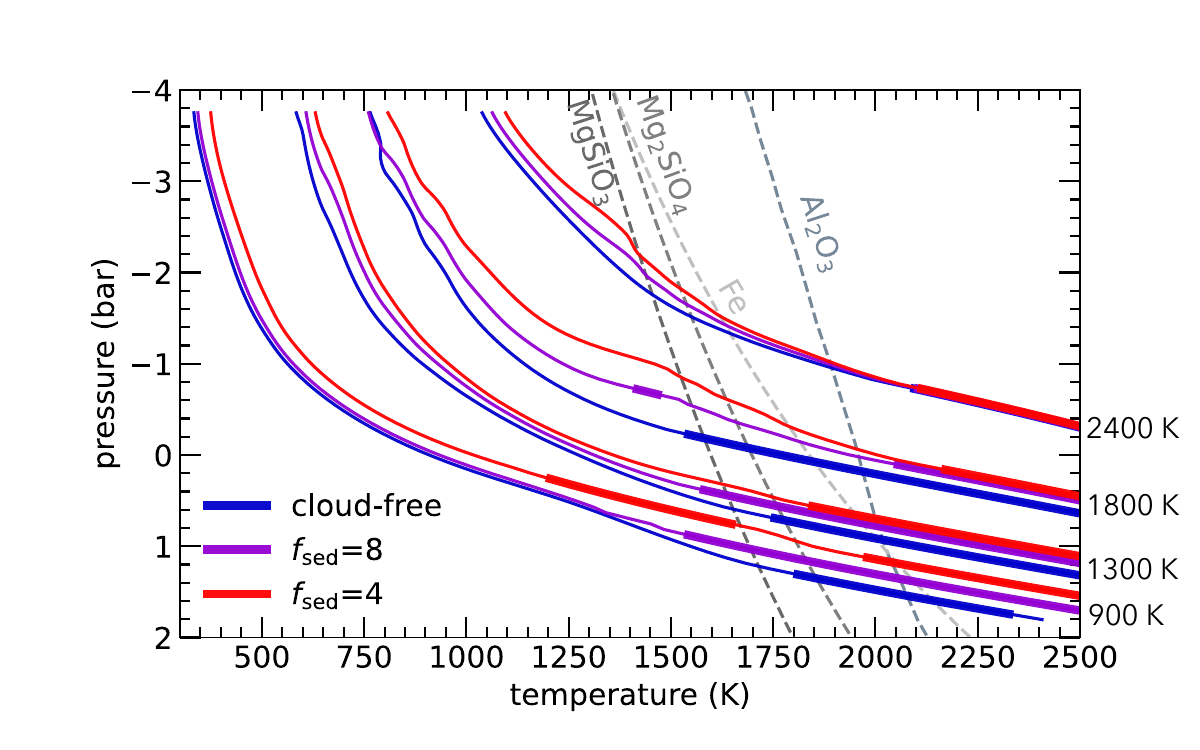}
    \caption{Pressure--temperature profiles of a selection of cloudy and cloud-free models from \teff=900--2400 K, with log g=4.5. Cloud-free models are shown as blue lines for comparison. Cloudy models with thinner clouds (\fsed=8) and thicker clouds (\fsed=4) are shown as purple and red lines. The thicker regions indicate convective parts of the atmosphere, while the thinner regions are radiative. Condensation curves for the four clouds included in this model grid are shown as dashed lines. Note that cloudy models are always warmer than a cloud-free model with the same effective temperature. }
    \label{fig:ptprof}
\end{figure}

\begin{figure}
    \includegraphics[width=3.5in]{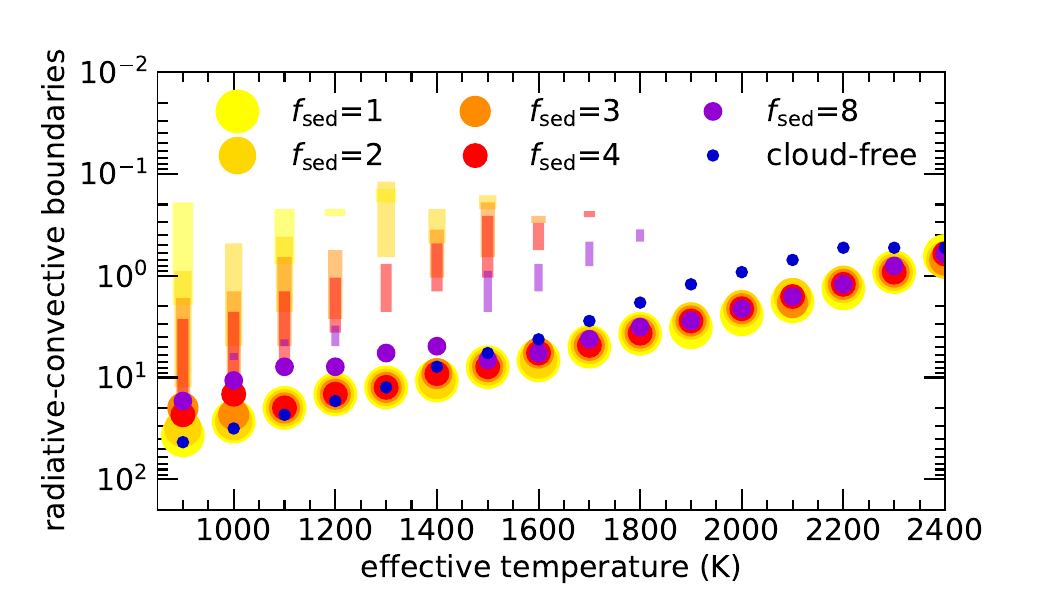}
    \caption{Radiative--convective boundaries for a selection of models. All models have log g=5.0. Solid circles show the deep radiative--convective boundary for each model; larger circles indicate cloudier models, as shown in the legend. The shaded rectangles show lower pressure detached convective zones. These are never present in the hottest models or cloud-free models, but start to appear at \teff$<$1900 K due to the additional opacity of clouds. }
    \label{fig:convzones}
\end{figure}

\begin{figure*}
    \centering
    \includegraphics[width=5in]{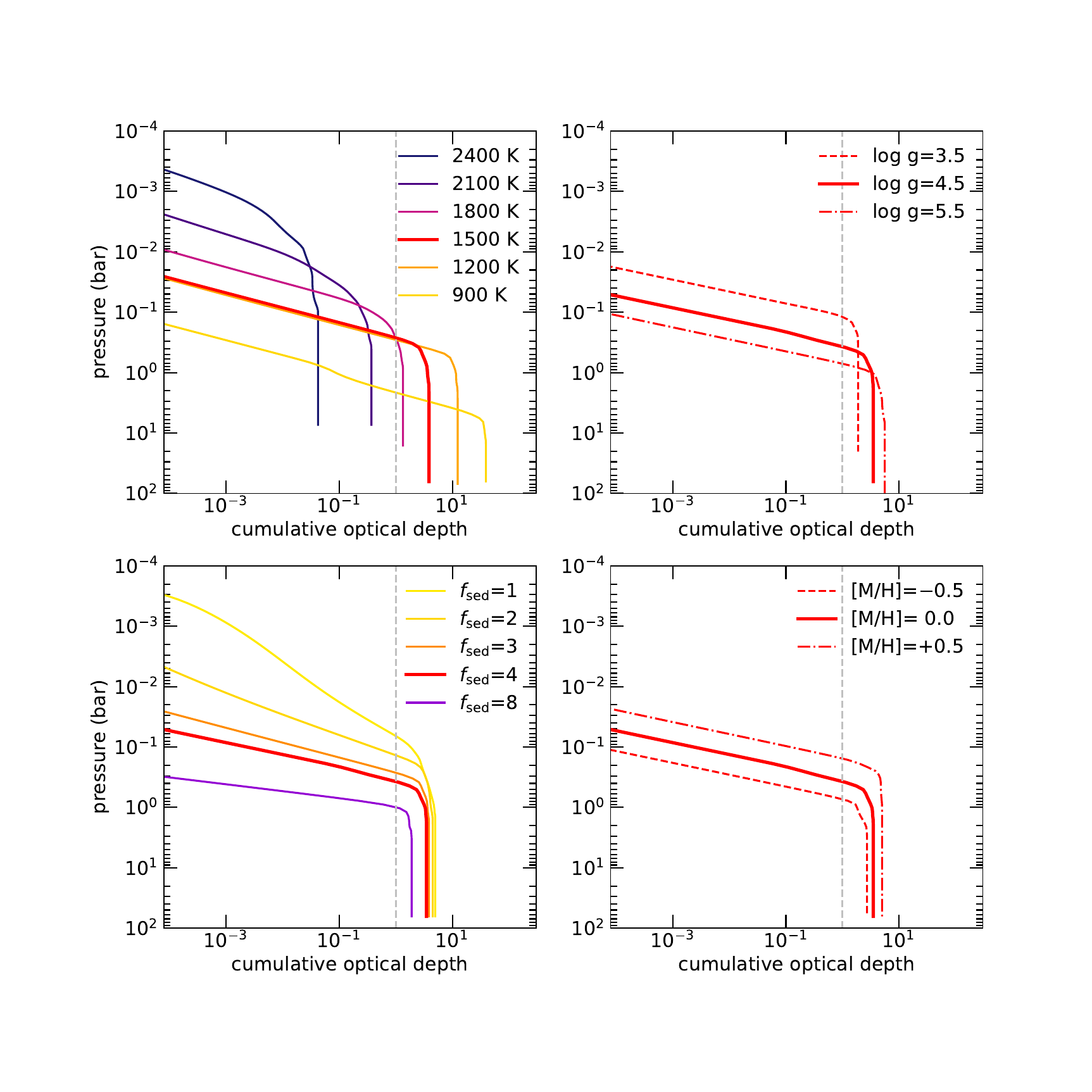}
    \caption{Cumulative optical depth as a function of each parameter in the Diamondback grid. Cumulative optical depth is calculated by summing $\tau$ at a wavelength of 1.1 $\mu$m) from the top of the atmosphere. $\tau$=1 is shown as a dashed vertical line to guide the eye. The top left panel shows that increasing effective temperature raises the cloud base height to lower pressures and decreases the overall optical depth of the cloud. The top right panel shows that decreasing gravity typically raises the cloud base to lower pressures. The bottom left panel shows that decreasing \fsed\ changes the vertical extent of the cloud. The bottom right shows that increasing metallicity creates a more optically thick cloud, raising the cloud base to lower pressures.   }
    \label{fig:cloudtaus}
\end{figure*}

\begin{figure*}[h]
    \centering
    \includegraphics[width=5.5in]{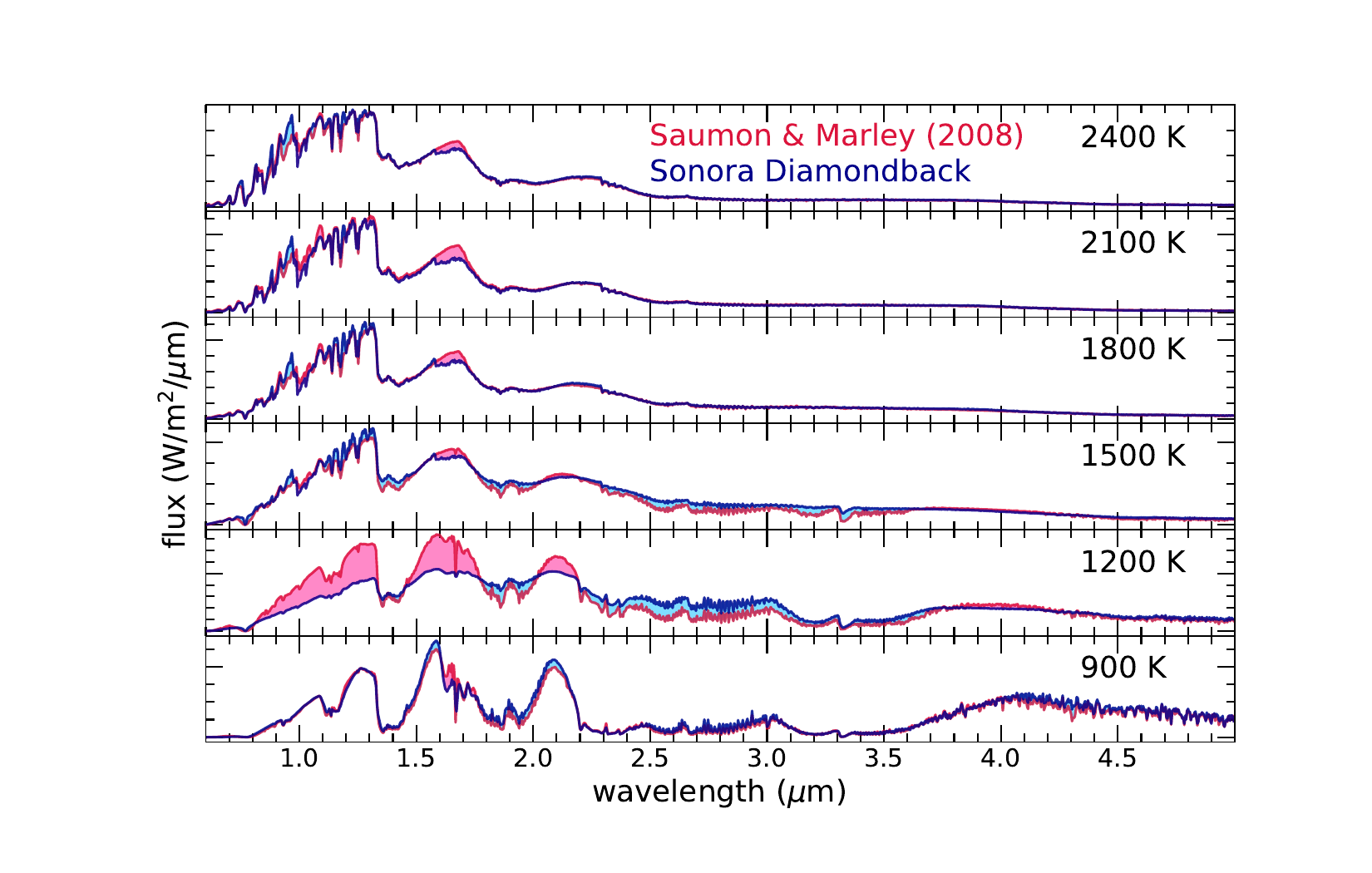}
    \caption{Sonora Diamondback models compared to the prior generation models from \citet{Saumon08}. All models have log g=5.0. The \citet{Saumon08} models have \fsed=2 while the Diamondback models have \fsed=3; the Diamondback models have slightly thicker clouds for a given \fsed\ due mostly to the addition of the MgSiO$_3$ cloud which was not included in \citet{Saumon08}. The area between the models is shaded blue where the Diamondback model is brighter and pink where the \citet{Saumon08} model is brighter. Changes can be seen at wavelengths affected by hydride opacities (e.g., FeH at 1.6 $\mu$m), oxide, and alkali opacities. The 1200 K model is somewhat of an outlier with a thicker cloud that reddens the spectrum.  }
    \label{fig:sm08}
\end{figure*}

\begin{figure*}
    \centering
    \includegraphics[width=5.5in]{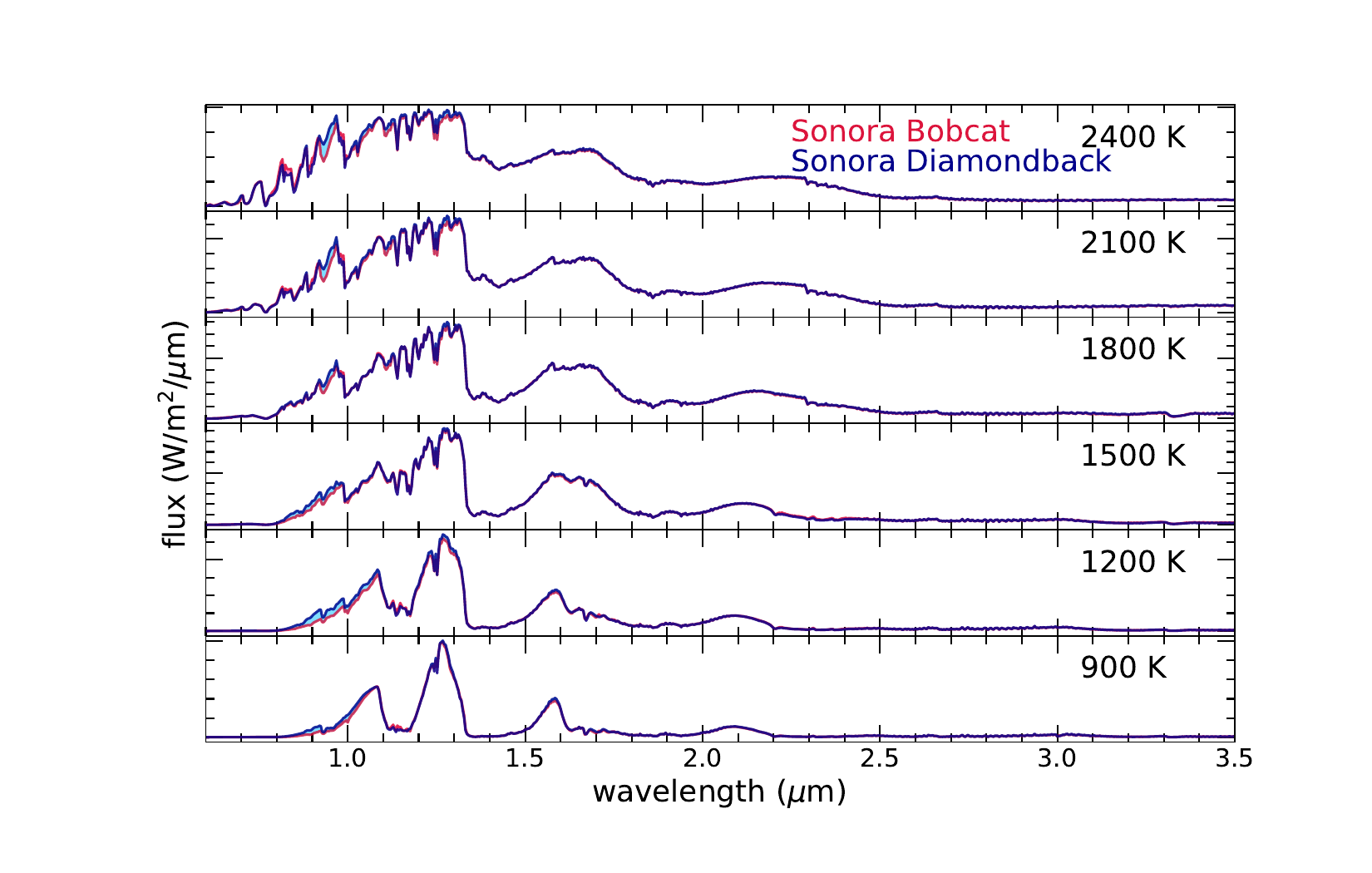}
    \caption{Sonora Diamondback models compared to the prior generation models from \citet{Marley21}. All models have log g=5.0 and are cloud-free. The small differences seen are attributable to differences in the line lists used, including the alkali lines and oxide/hydride features.  }
    \label{fig:bobcat}
\end{figure*}

\subsection{P--T Profiles}

We calculate the temperature structure of the atmosphere self-consistently with the condensation of clouds. As in previous modeling of L dwarfs, the clouds have a `blanketing' effect, warming the atmosphere throughout. This warming is quite significant: the cloudy models can be hundreds of degrees hotter at a given pressure than a cloud-free model with the same luminosity. The opacity of clouds can also change the locations of convective regions of the atmosphere, often generating convective regions in layers that would be radiative in a cloud-free atmosphere.

We show examples of P--T profiles in Figure \ref{fig:ptprof}, at 900, 1300, 1800, and 2400 K (log g=4.5). We also show condensation curves for the four clouds included in our modeling (MgSiO$_3$, Mg$_2$SiO$_4$, Fe, and Al$_2$O$_3$), which show where the vapor pressure of the limiting species is equal to the saturation vapor pressure for solar metallicity. The thicker lines indicate convective regions of the atmosphere while the thinner lines indicate radiative regions. At 2400 K, the clouds form at low pressures and are relatively optically thin, leading to small changes in the P--T profile and no change to the radiative--convective boundary. In contrast, at 900 K, the clouds are quite optically thick and the atmospheric structure is more complex, with a significant convective region at lower pressures now present in the \fsed=4 atmosphere. 

This behavior is typical for our model grid. We show the radiative--convective boundaries for a broader range of models in Figure \ref{fig:convzones}. The circles on this plot show the location of the deep radiative--convective boundary. For a given surface gravity (here, log g=5.0), the radiative--convective boundary becomes deeper with cooler effective temperature, increasing from 0.6 bar at 2400 K to 40 bar at 900 K. Cloud-free models never have lower pressure detached convective regions in this temperature range, while cloudy models begin to have these detached convective regions at 1600--1800 K. These regions grow with decreasing temperature.

\subsection{Cloud optical depth}

The thickness of clouds in our models is strongly dependent on \teff, log g, \fsed, and metallicity. In Figure \ref{fig:cloudtaus} we show how the optical depth (summed from the top of the atmosphere at a wavelength of 1.1 \micron) varies with each of these properties. We find that clouds become more optically thick with decreasing effective temperature; the warmest models at this surface gravity with a cumulative optical depth of 1 are 1800 K. However, for cooler objects the cloud is deeper in the atmosphere, eventually sinking below the near-infrared photosphere. Thus while the coldest temperature models always have the thickest clouds, the clouds may peak in their spectral impact at a higher \teff. 

Surface gravity and metallicity both change the pressure of the cloud base (typically to lower pressures for lower gravities and higher metallicities), and can also change the total cumulative depth; these properties typically do not alter the vertical structure of the cloud (i.e., the slope of the optical depth with pressure is the same). In contrast, \fsed\ changes the vertical structure without changing the cloud base pressure significantly. Lower \fsed\ values create more vertically lofted clouds (with smaller particles), while higher \fsed\ values create more compact clouds (with larger particles). 

\begin{figure*}
    \centering
    \includegraphics[width=6.5in]{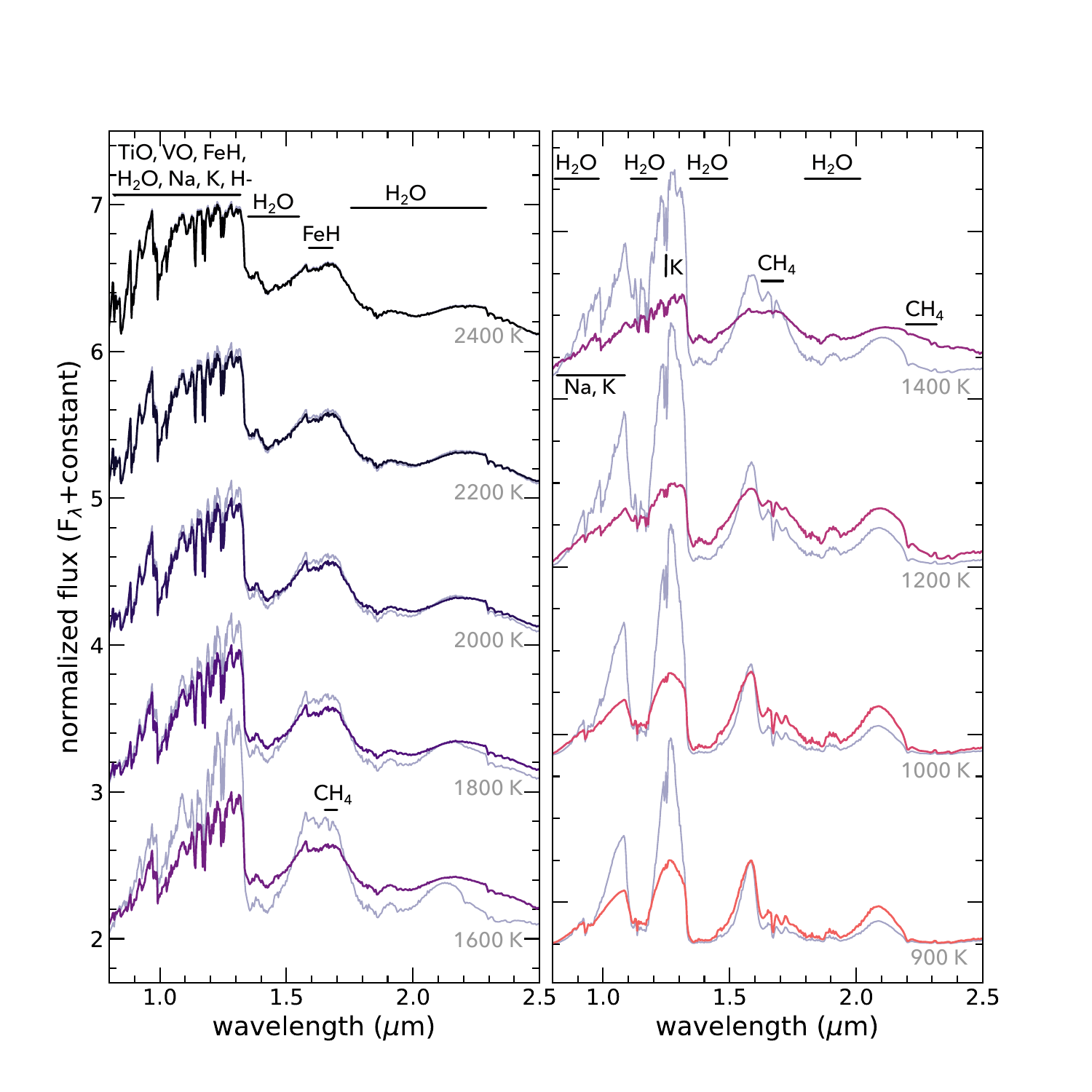}
    \caption{Sonora Diamondback model spectra in the near-infrared. Models are shown for a range of \teff\ from 2400 (top left) to 900 K (bottom right). All models shown have log g=5.0. Cloudy (\fsed=4) spectra are shown as colored lines; cloud-free models are shown as gray lines for each \teff. Key molecular features are labeled. }
    \label{fig:spectra_fullnearir}
\end{figure*}

\begin{figure*}
    \centering
    \includegraphics[width=5.5in]{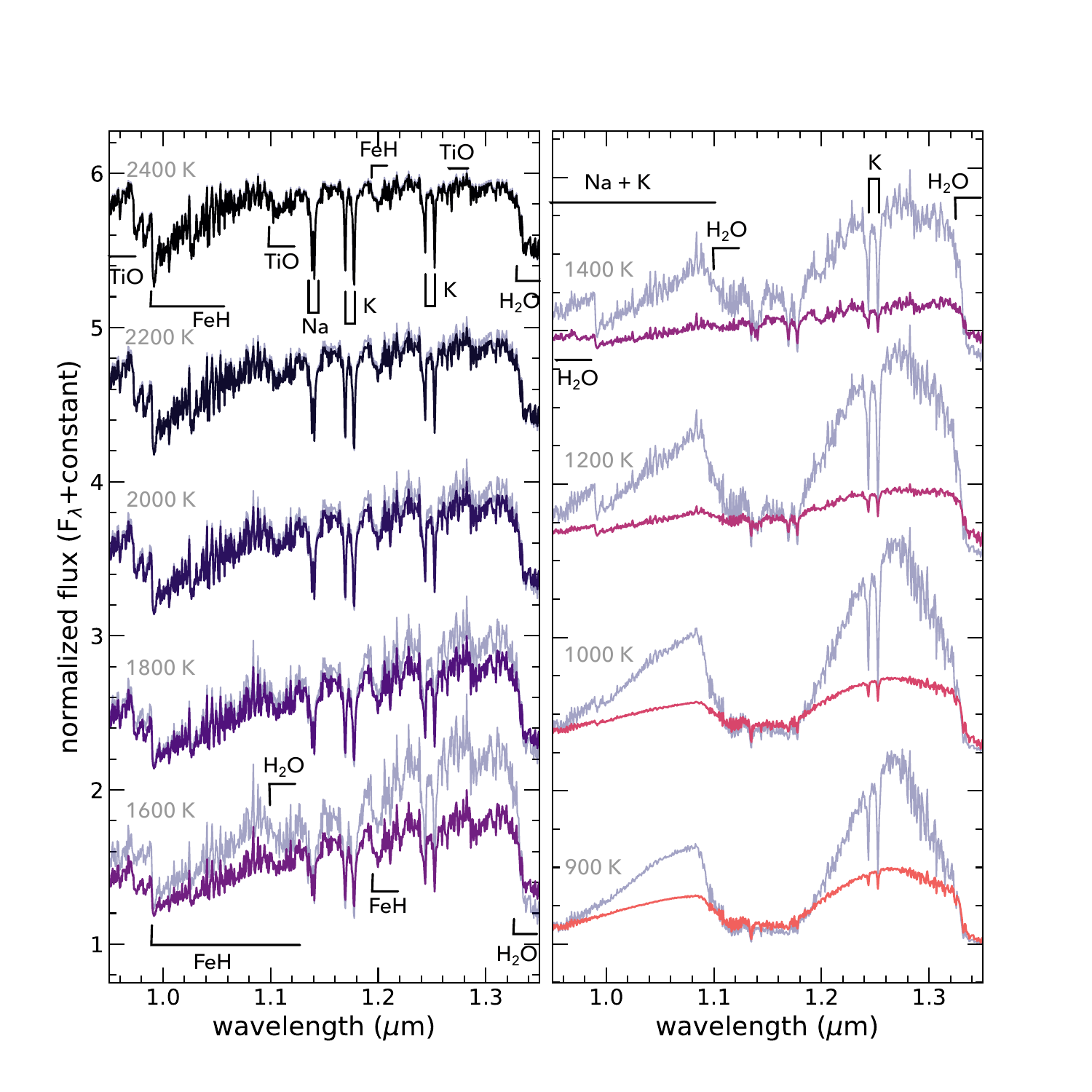}
    \caption{Sonora Diamondback model spectra in Y and J bands (0.9-1.35 \micron). Models are shown for a range of \teff\ from 2400 (top left) to 900 K (bottom right). All models shown have log g=5.0. Cloudy (\fsed=4) spectra are shown as colored lines; cloud-free models are shown as gray lines for each \teff. Key molecular features are labeled. }
    \label{fig:spectra_zoomnearir}
\end{figure*}

\begin{figure*}
    \centering
    \includegraphics[width=6.5in]{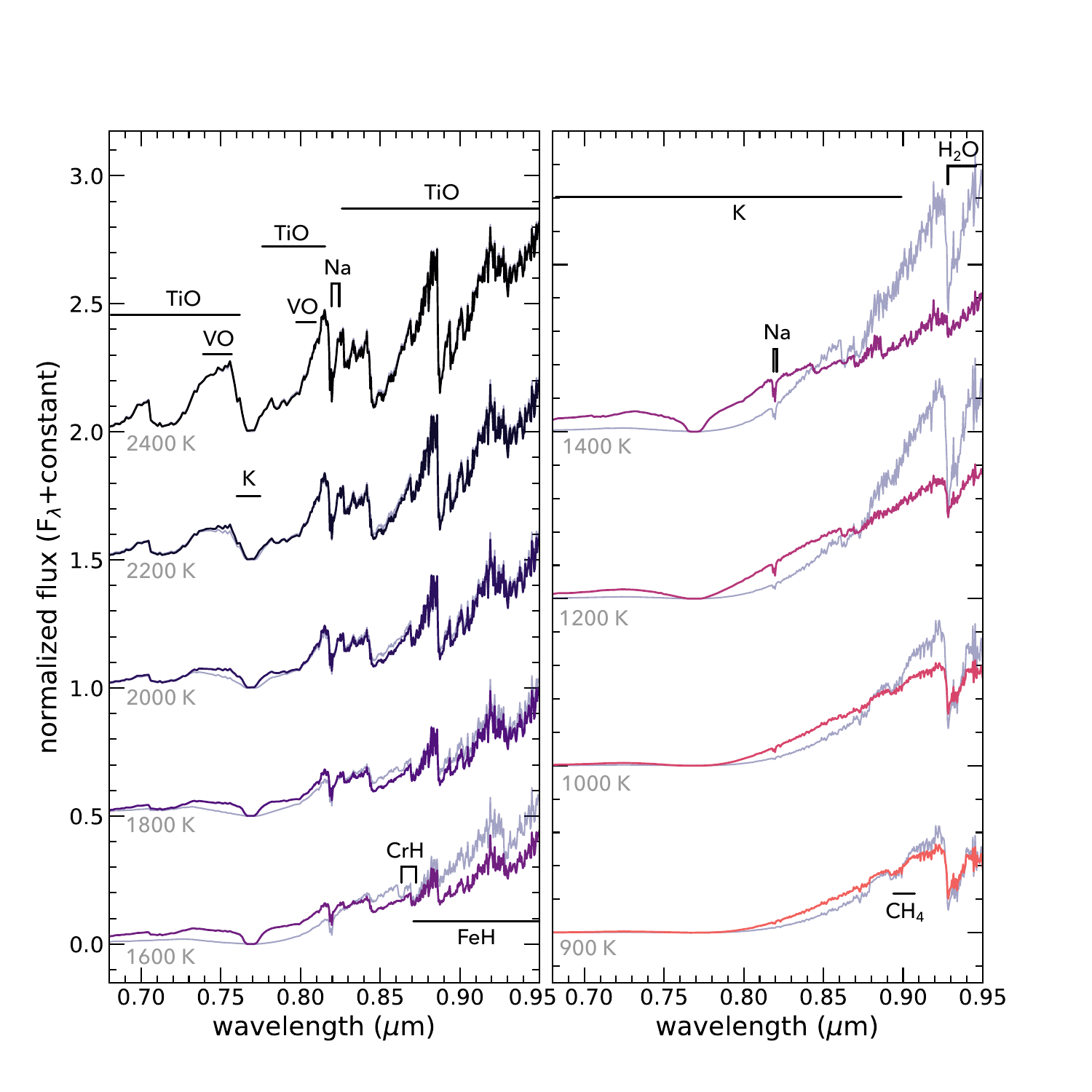}
    \caption{Sonora Diamondback model spectra in the red optical. Models are shown for a range of \teff\ from 2400 (top left) to 900 K (bottom right). All models shown have log g=5.0. Cloudy (\fsed=4) spectra are shown as colored lines; cloud-free models are shown as gray lines for each \teff. Key molecular features are labeled. }
    \label{fig:spectra_opt}
\end{figure*}

\begin{figure*}
    \centering
    \includegraphics[width=6.5in]{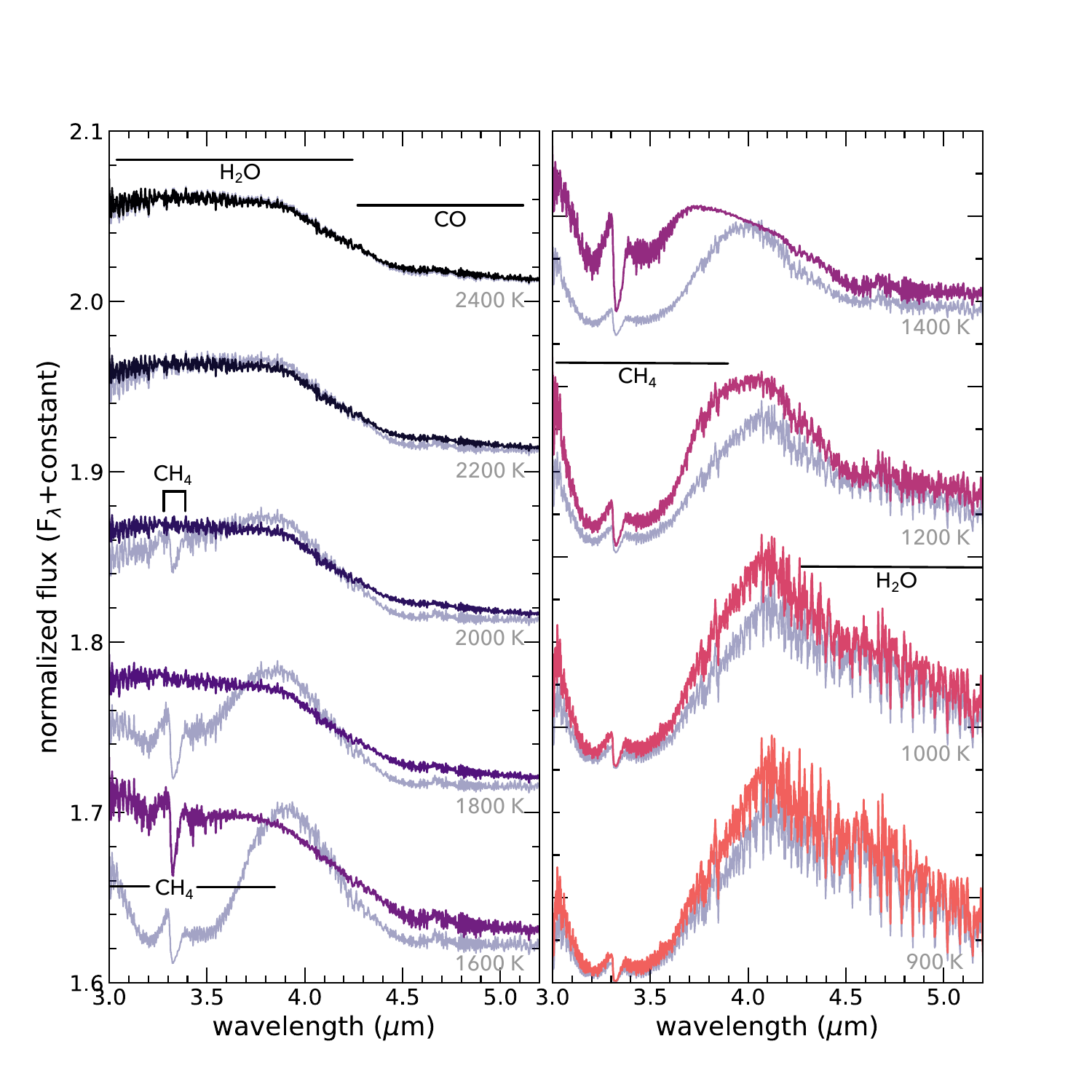}
    \caption{Sonora Diamondback model spectra in the mid-infrared (3--5.2 \micron). Models are shown for a range of \teff\ from 2400 (top left) to 900 K (bottom right). All models shown have log g=5.0. Cloudy (\fsed=4) spectra are shown as colored lines; cloud-free models are shown as gray lines for each \teff. Key molecular features are labeled. }
    \label{fig:spectra_midir}
\end{figure*}

\begin{figure*}
    \centering
    \includegraphics[width=6.5in]{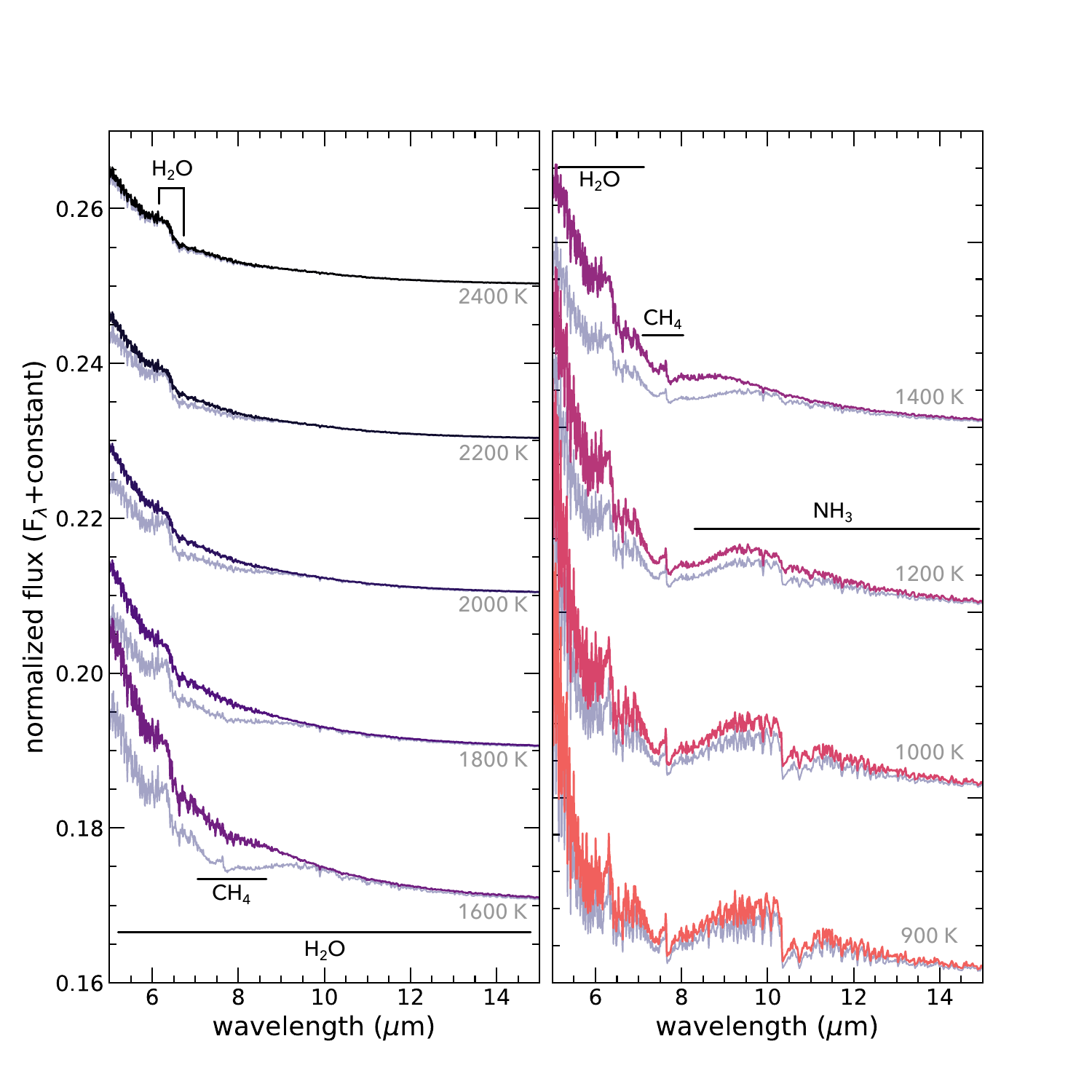}
    \caption{Sonora Diamondback model spectra in the thermal infrared (5.5-15 \micron). Models are shown for a range of \teff\ from 2400 (top left) to 900 K (bottom right). All models shown have log g=5.0. Cloudy (\fsed=4) spectra are shown as colored lines; cloud-free models are shown as gray lines for each \teff. Key molecular features are labeled. }
    \label{fig:spectra_thermalir}
\end{figure*}

\begin{figure*}
    \centering
    \includegraphics[width=7in]{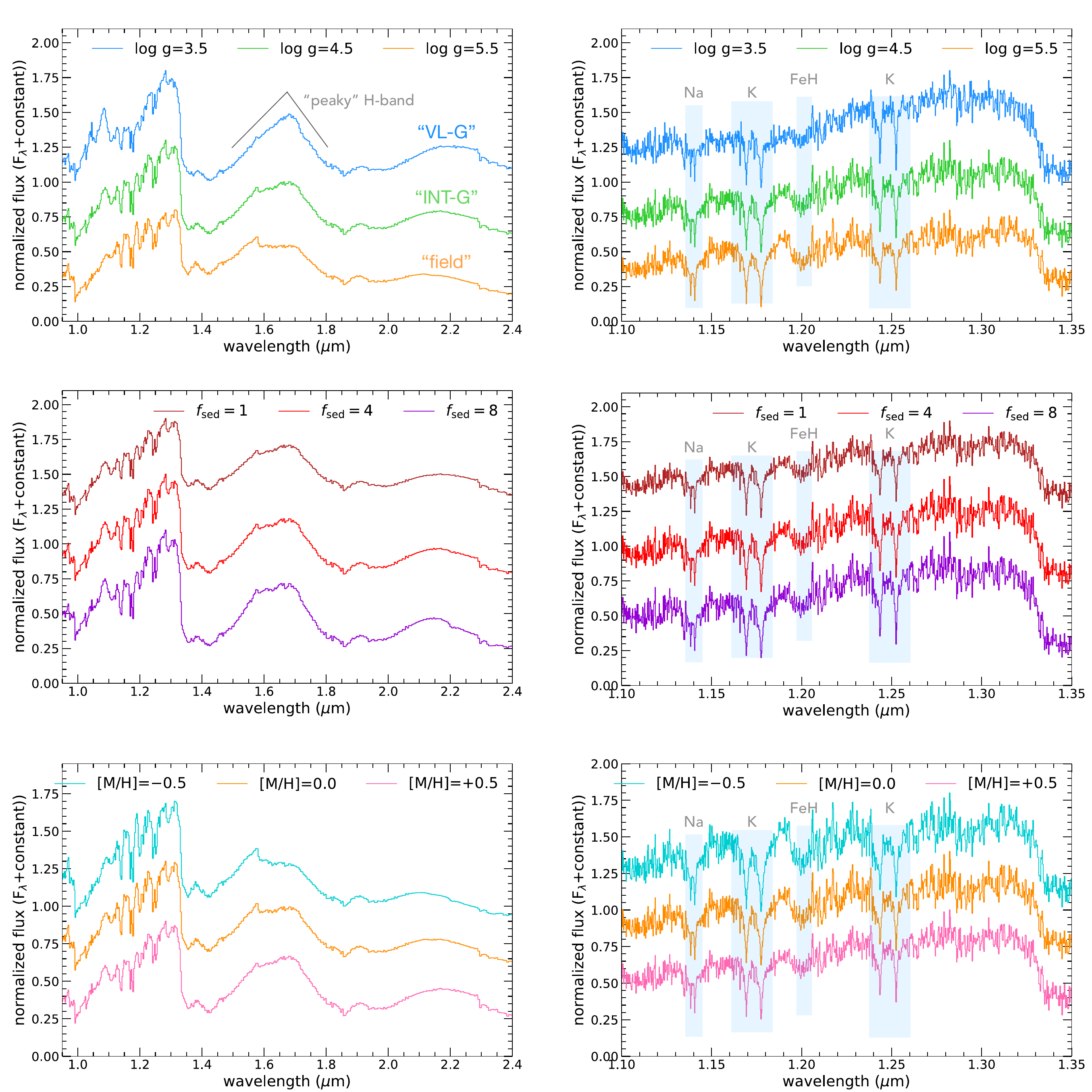}
    \caption{The effect of surface gravity, cloud thickness, and metallicity on spectra. All models have \teff=1500 K. The left panels show the broad near-infrared spectra, while the right panels zoom into the gravity-sensitive features within J band. The top panel shows the effect of surface gravity for models with solar metallicity and \fsed=3; spectral features that have been identified to empirically correlate with surface gravity are labeled. The middle panel shows the effect of cloud sedimentation efficiency for models with solar metallicity and log g=4.5. The bottom panels show the effect of metallicity for models with log g=4.5 and \fsed=4. }
    \label{fig:spectra_compiled}
\end{figure*}

\subsection{Model Spectra}

Model spectra are calculated using the \texttt{disort}-based radiative transfer code presented in \citet{Stamnes88, Morley15, Stamnes17}. This radiative transfer model provides very good agreement with that used in the \citet{Marley21} Sonora Bobcat models and is the same as is used in the Sonora Cholla models \citep{Karalidi21}. The model includes multiple-scattering of clouds using a Henyey-Greenstein phase function and the discrete-ordinates method to solve the radiative transfer equations. 

We note that this \texttt{disort}-based code differs in its treatment of scattering from the \citet{Toon89} approach used within our 1D self-consistent climate model for calculating the P--T profiles. 
Recently a new radiative transfer method \citep {Rooney23a, Rooney23b} pointed out differences between radiative transfer methods which utilize a hemispheric mean phase function \cite[e.g.,][]{Toon89}, and radiative transfer methods which use a Henyey-Greinstein phase function (e.g., spherical harmonics or \texttt{disort}). These differences are most prominent for moderately  scattering (single scattering albedo $\omega_0\sim0.5-0.8$), forward scattering ($g_0>$0.4) cloud particles, as studied in detail in \citet{Rooney23a, Rooney23b}. 
Unfortunately the silicate clouds presented here do sometimes fall in this parameter space, and we thus see some discrepancies between the low-resolution spectra from the climate model and the high-resolution post-processed spectra. \citet{Rooney23b} devised a two-term spherical harmonics routine that could be implemented within our climate framework to rectify these discrepencies in future work.\footnote{\citet{Rooney23b} is still under review and as such the code is still being actively developed. Therefore, it is not possible to include in this iteration of the model grid.}  

We present our model spectra in a series of Figures aimed at showing the changes from prior generations of models, showing and labeling the key spectral features at each wavelength, and demonstrating how clouds, surface gravity, and metallicity affect the spectra. 

\subsection{Changes from Saumon \& Marley 2008 and Sonora Bobcat}

Figure \ref{fig:sm08} shows how the Sonora Diamondback models differ from models with the same temperatures and surface gravities from the \citet{Saumon08} grid. The biggest difference is that the clouds are thicker (due to the inclusion of the MgSiO$_3$ cloud) for all Diamondback models. For this reason, we show \fsed=3 clouds for the Diamondback models and \fsed=2 clouds for the \citet{Saumon08} models. The other changes in the spectra are due to changes in the molecular line lists used, including Na, K, TiO, and VO in the optical and near-IR; FeH at 1.5-1.7 \micron; and CH$_4$ at 1.6-1.7 \micron. The 1200 K model has notably thicker clouds in the Sonora Diamondback models at this \fsed\ pair, which reddens the 1-2.25 \micron\ region and increases the flux from 2.25-3.1\micron, but this is not necessarily representative of the whole grid.

Figure \ref{fig:bobcat} shows how the Sonora Diamondback \emph{cloud-free} models differ from the Sonora Bobcat models. These changes are small because the ingredients in these models are nearly the same. The updates since the \citet{Marley21} Bobcat models include: changes in the alkali, H$_2$O, and CH$_4$ line lists and opacities, as well as the addition of new species: Fe, H$_2$, H$_3^+$, LiF, and LiH.
The existing opacities and chemistry have also been re-gridded or re-calculated on a 1460-P/T point grid instead of a 1060-P/T point grid. The new P/T point grid aims to more finely capture the areas where the opacity changes quickly, which includes where water condenses (colder than the models here) and where CO and CH$_4$ change quickly in abundance as a function of temperature. The updated set of k-coefficients can be found on Zenodo \citep{zenodo-k}, accompanied by the corresponding set of opacities calculated at higher resolution \citep{zenodo-opac}.

\subsection{Cloudy model spectra}

Figures \ref{fig:spectra_fullnearir}, \ref{fig:spectra_zoomnearir}, \ref{fig:spectra_opt}, \ref{fig:spectra_midir}, and \ref{fig:spectra_thermalir} show cloud-free and cloudy (\fsed=4) spectra from the Sonora Diamondback at 900--2400 K. At \teff=2200--2400 K, the clouds are optically thin and the cloudy and cloud-free spectra are nearly identical. Their spectra are shaped by atomic features (Na, K) and metal hydrides and oxides (FeH, TiO, VO) in the optical and near-infrared, and by H$_2$O and CO in the mid- and thermal infrared. For cooler models, the oxides and hydrides weaken in strength and eventually disappear. For example, FeH has a strong feature at 0.98 \micron, best shown in Figure \ref{fig:spectra_zoomnearir}; this feature is weak in amplitude by 1400 K, and is no longer seen by eye in the spectra by 900 K. 

Clouds have a number of major effects on spectra from the optical through thermal infrared. First, the opacity of clouds is relatively gray in the near-infrared, adding significant opacity at wavelengths where, without clouds, the atmosphere is relatively clear, allowing us to see deeply within the atmosphere. This cloud opacity effect reddens near-infrared spectra, decreasing the flux especially from 1--1.4 \micron. Because of the opacity of clouds, the strength of many atomic and molecular features decreases compared to cloud-free models, despite no appreciable change in the abundance of those species. This is perhaps seen best in Figure \ref{fig:spectra_zoomnearir}, for the K features at 1.24-1.26 \micron. 

Lastly, the clouds warm the P/T profile (see Figure \ref{fig:ptprof}). This warming of the profile shifts the emergence of CH$_4$ as a function of temperature by 300--400 K, and shifts the strength of the CO and CH$_4$ features in 1200--2000 K models. These shifts are best seen in Figure \ref{fig:spectra_midir}. Cloud-free models have cooler P/T profiles for a given \teff; CH$_4$ is first seen at $\sim$3.3 \micron\ (the Q branch of the fundamental $\nu_3$ methane band) around \teff=2000 K. For cloudy models, since their P/T profiles are warmed by the clouds, CH$_4$ is instead first seen at \teff=1600-1700 K. The difference in these models, which assume chemical equilibrium, is maximized around 1400--1800 K; at cooler temperatures, the cloudy and cloud-free spectra in the thermal infrared are much more similar in shape, since both are dominated by CH$_4$, H$_2$O, and, NH$_3$. 

Figure \ref{fig:spectra_compiled} shows, for a single \teff\ near the center of our grid (1500 K), how surface gravity, cloud thickness, and metallicity impact key diagnostic regions of the spectra. We model these plots after the spectra figures of \citet{Allers13}, who developed some of the key spectral features and corresponding spectral indices for low gravity L dwarfs. We highlight some of the regions of the spectrum that \citet{Allers13} empirically correlate with low surface gravity. This includes the `peaky' H band, where lower gravity models have a more triangular spectrum from 1.6--1.8 \micron. We reproduce this in the lowest gravity model shown here, while the higher gravity objects have a broader H band peak with an FeH feature present. In the high resolution J band spectra shown on the right of Figure \ref{fig:spectra_compiled}, we highlight Na, K, and FeH features. These features are \emph{weaker} in low-gravity objects \citep{Allers13}. We see in contrast that---keeping \fsed\ equal---these features actually become \emph{stronger} at low surface gravity in our models. However, they become weaker with thicker clouds (see the center panel here). We suggest that in observed brown dwarf spectra, the impact of thicker clouds in low-gravity objects ends up weakening these features, outweighing the opposing effect of changing only surface gravity. 

In Figure \ref{fig:spectra_compiled} we also show the impact of metallicity, which also changes the shape of the spectrum. Here we see the most pronounced, distinctive changes in the K band region of the spectrum, from 2--2.4 \micron, where the higher metallicity model has a more distinctive CO feature at 2.3 \micron\ while the lower metallicity model has a smoother spectrum there, shaped by collision-induced opacity of H$_2$.

\subsection{Thermal evolution models}

\begin{figure*}
    \includegraphics[width=7.5in]{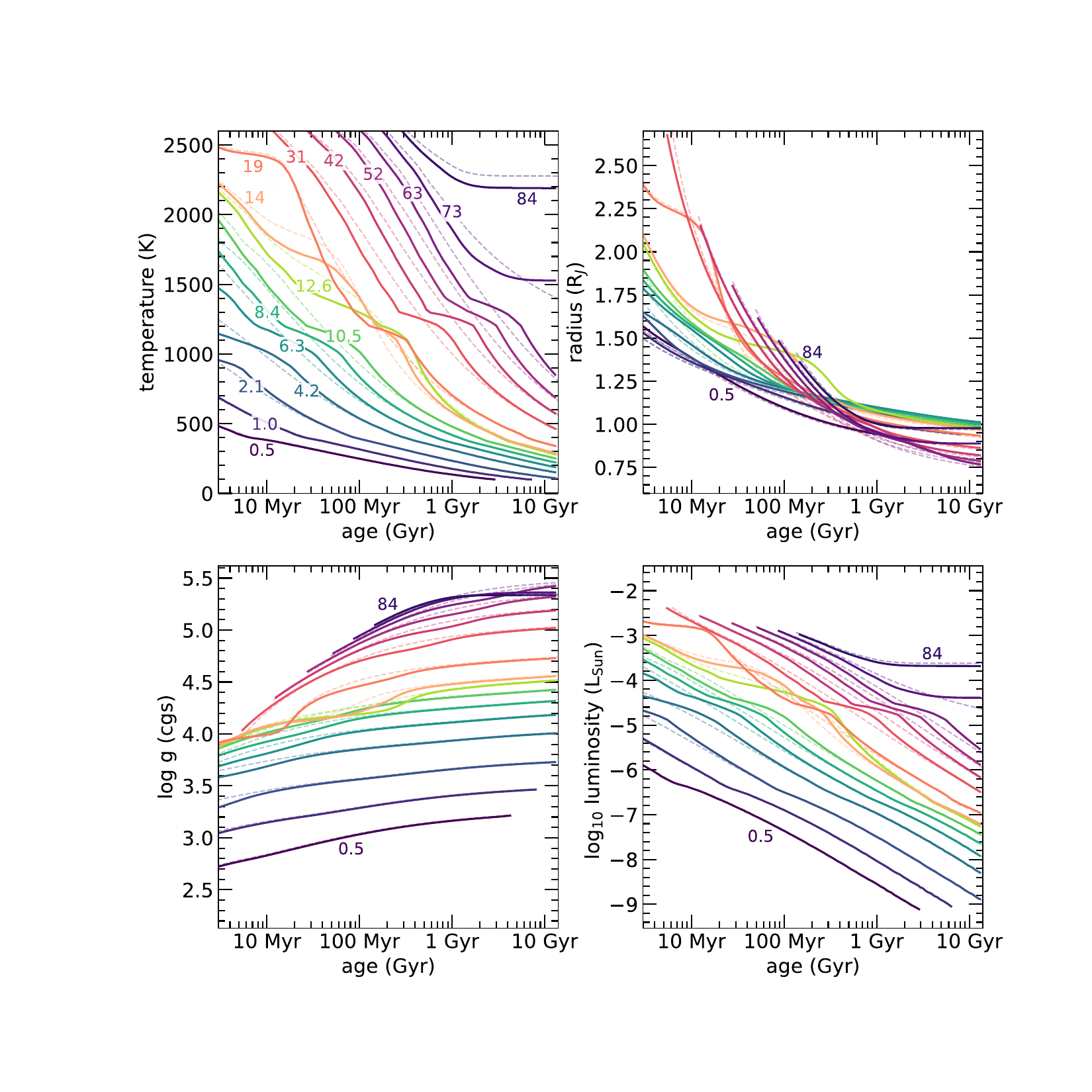}
    \caption{Thermal evolution models for cloud-free and hybrid models at solar metallicity. Cloud-free models are shown as dashed lines and hybrid models are shown as solid lines. The clouds in the hybrid models clear at 1300 K, changing the evolution. The top left panel shows temperature vs age. Atmospheres cool slightly faster until the transition, and then stall at $\sim$1200 K when the clouds clear, remaining slightly warmer than the cloud-free models of the same age. Radius vs. age (top right) shows that objects shrink as they age, from 1.5-2.5 R$_J$ at 3 Myr to 0.75-1 R$_J$ at 10 Gyr. log g vs. age (bottom left) shows that objects increase in surface gravity as they age, typically by roughly 0.5 dex. Luminosity vs. age (bottom right) looks similar to \teff\ vs. age, and shows that an objects luminosity evolution is shaped by the emergency and disappearance of clouds.  }
    \label{fig:evolution}
\end{figure*}

\begin{figure*}
    \centering
    \includegraphics[width=7.5in]{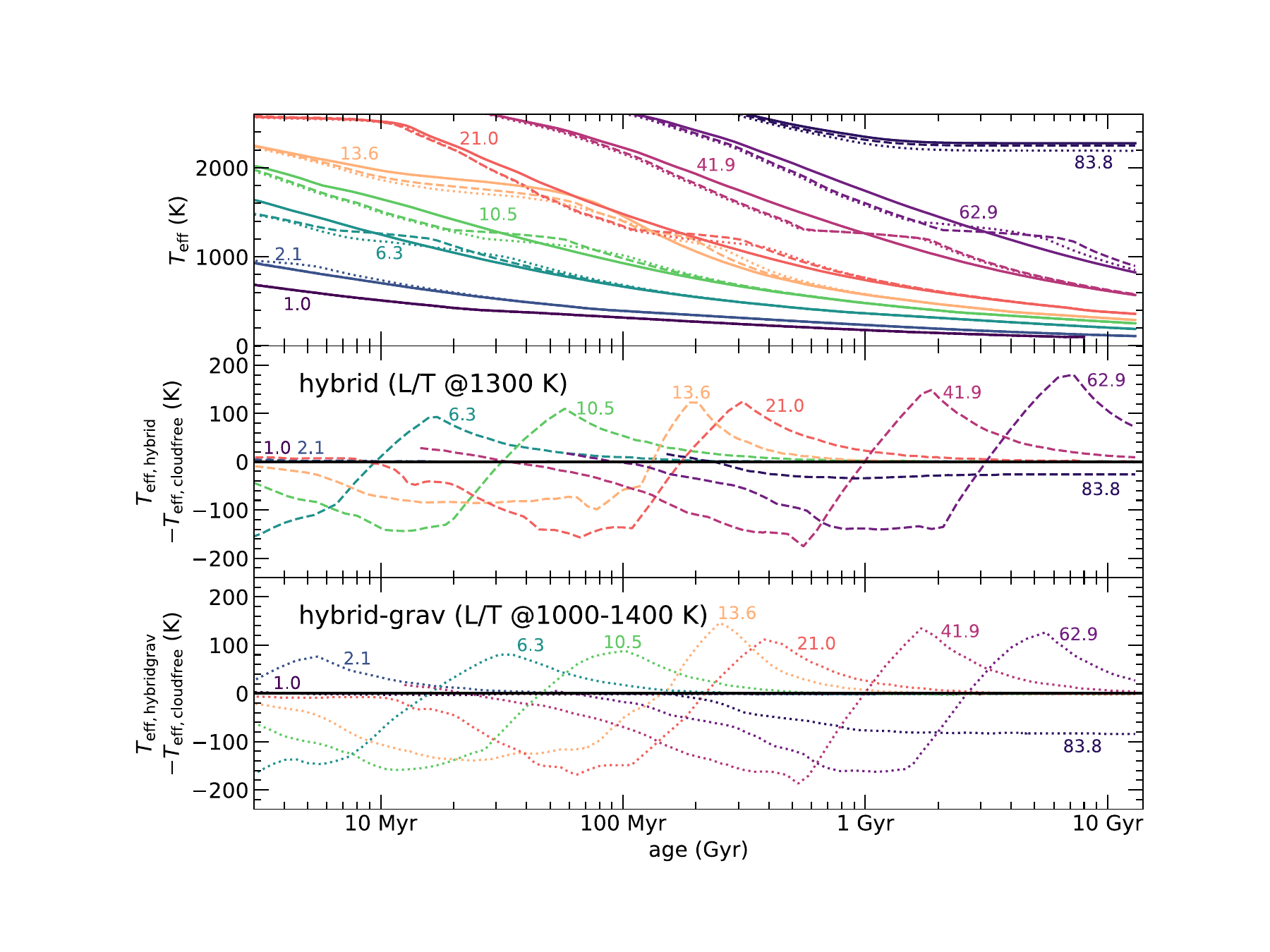}
    \caption{Temperature evolution for cloud-free and hybrid models at solar metallicity. Cloud-free models are shown as solid lines; hybrid models are shown as dashed lines; hybrid-grav models are shows as dotted lines. The clouds in the hybrid models clear at 1300 K, while in the hybrid-grav models they clear at 1000-1400 K (dependent on surface gravity). The top panel shows temperature vs age. The bottom two panels show the temperature difference between the hybrid and hybrid-grav models and cloud-free models. Clouds can change the effective temperature at a given age by 100--200 K.}
    \label{fig:evolution_clouds}
\end{figure*}

\begin{figure*}
    \centering
    \includegraphics[width=6in]{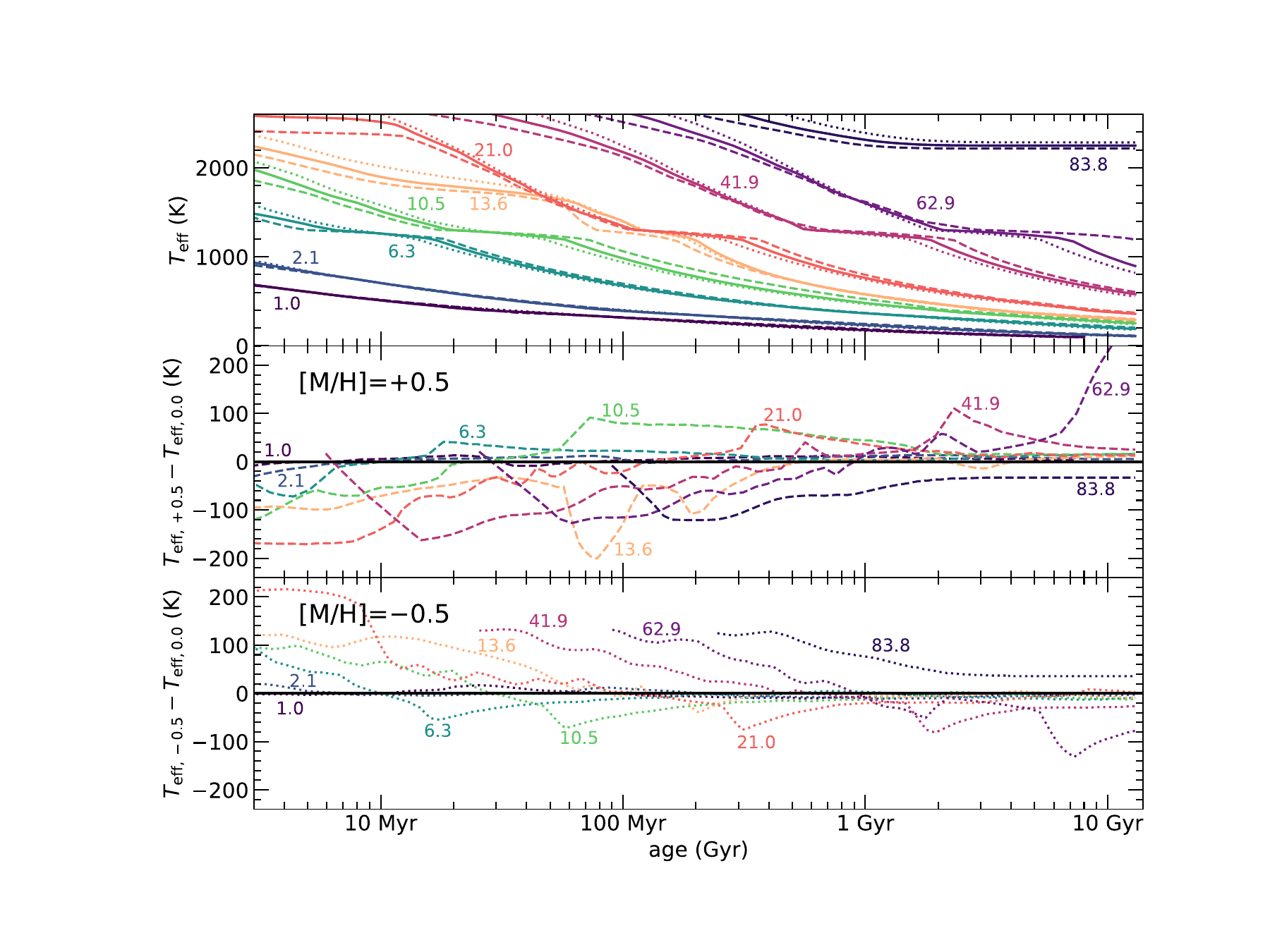}
    \caption{Temperature evolution for hybrid models at three metallicities. Solar metallicity models are shown as solid lines; [M/H]=$+0.5$ models are shown as dashed lines; [M/H]=$-0.5$ models are shows as dotted lines. The top panel shows temperature vs age. The bottom two panels show the temperature difference between the [M/H]=$+0.5$ and [M/H]=$-0.5$ models and solar metallicity models. Metallicity can change the effective temperature at a given age by 100--200 K.}
    \label{fig:evolution_met}
\end{figure*}

As described above (Section \ref{sec:evolution}), we used the atmosphere models as boundary conditions for calculating the thermal evolution of brown dwarfs and giant planets. We considered both cloud-free evolution and `hybrid' evolution where objects start cloudy and then transition to cloud-free below the L/T transition. We build on prior work by considering both a single-temperature transition from cloudy to cloud-free (`hybrid') and a gravity-dependent transition (`hybrid-grav') where the temperature of the cloudy-to-clear transition is lower for lower gravities. Our evolution modeling results are summarized in Figures \ref{fig:evolution}, \ref{fig:evolution_clouds}, and \ref{fig:evolution_met}. 

Figure \ref{fig:evolution} shows the temperature, radius, log g, and luminosity of our evolution models as a function of age. Cloud-free models are shown as dashed, transparent lines, and `hybrid' models (with an `L/T' transition at 1300 K) are shown as solid, opaque lines. The top left and bottom right panels show the temperature and luminosity evolution respectively. During the time when the hybrid models are cloudy (\teff$>$1300K), the models cool faster; when the clouds clear, cooling stalls. Eventually the cooling tracks of the hybrid and cloud-free models meet again. 

As seen in prior studies \citep[e.g.,][]{Saumon08}, deuterium fusion at different rates and times for 11--20 M$_J$ models can cause objects with the same mass to overlap in their cooling tracks, especially at young to intermediate ages ($\sim$30--500 Myr). For example, for the models shown in Figure \ref{fig:evolution}, at 200 Myr, the hybrid 19 M$_J$, 14 M$_J$, and 12.6 M$_J$ models all have the same \teff, $\sim$1200 K. The masses of objects inferred from evolution models just above the deuterium-burning limit are therefore likely to be more uncertain than for other mass ranges.

The radius evolution is shown in the upper right panel of Figure \ref{fig:evolution}; all models shrink in radius over time, from over 1.5 R$_J$ at 3 Myr to 0.75--1.1 R$_J$ at 10 Gyr. Deuterium fusion can stall the radius evolution in objects, especially just above the deuterium-burning mass limit. Likewise, the surface gravity evolution is shown in the lower left panel of Figure \ref{fig:evolution}. As expected, surface gravity increases slightly as the radius shrinks over time. 

We highlight how the clouds and metallicity impact the temperature evolution in Figures \ref{fig:evolution_clouds} and \ref{fig:evolution_met}. In Figure \ref{fig:evolution_clouds}, we compare cloud-free evolution models to `hybrid' (transition to cloud-free at 1300 K) and `hybrid-grav' (transition to cloud-free at 1000-1400 K) models. The bottom two panels show the difference in \teff\ between the cloud-free evolution and hybrid evolution. The effect of clouds on the evolution can change the \teff\ at a given age by up to 100--200 K, depending on the mass of the object. Likewise in Figure \ref{fig:evolution_met} we show `hybrid' evolution for three different metallicities. The metallicity can also change the \teff\ at a given age by 100-200 K. 

\subsubsection{The deuterium-burning limit}

We can determine the deuterium burning limit in our evolution models and compare to prior models; we define our limit (somewhat arbitrarily) to be the mass above which an object burns more than 50\% of its initial deuterium. Within our model grid, we find slightly lower deuterium-burning limits for all hybrid models by 0.1 M$_J$ compared to cloud-free models. Metallicity has a larger impact on the deuterium-burning limit. We find the following limits: [M/H]=0.0: 12.05 M$_J$ (hybrid) and 12.16 (cloud-free), [M/H]=$-$0.5: 12.47 M$_J$ (hybrid) and 12.68 (cloud-free), [M/H]=$+$0.5: 11.21 M$_J$ (hybrid) and 11.32 (cloud-free). These limits are lower by $\sim$1 M$_J$ compared to models from \citet{Spiegel11}. 

We can also compare to models from \citet{Saumon08} and Sonora Bobcat \citep{Marley21}, who report the deuterium-burning limit as the mass above which an object burns 90\% of its deuterium within 10 Gyr: for cloud-free, solar metallicity models we find the same limit of 12.9 M$_J$ for Sonora Diamondback models as Sonora Bobcat models, very slightly lower than the 13.1 M$_J$ reported in \citet{Saumon08}. For hybrid solar metallicity models, we find a limit of 12.7 M$_J$, close to the 12.4 reported in \citet{Saumon08} for fully-cloudy tracks. We note that these deuterium-burning limits could be potentially tested with high fidelity JWST or ground-based spectra of cool brown dwarfs \citep{Morley19, Molliere19}. 

\subsubsection{The hydrogen-burning limit}

We can also determine the hydrogen burning limit and compare to prior models. We calculated the hydrogen-burning minimum mass (HBMM) by calculating the minimum mass for which the H-caused nuclear luminosity becomes independent of age within 10 Gyr. 

We find the following limits: 

\begin{itemize}
    \item solar metallicity, hybrid: 70.2 M$_J$ (=0.067 M$_{\rm Sun}$) 
    \item solar metallicity, cloud-free: 76.5 M$_J$ (=0.073 M$_{\rm Sun}$)
    \item +0.5, hybrid: 66.0 M$_J$ (=0.063 M$_{\rm Sun}$)
    \item +0.5, cloud-free: 70.2 M$_J$ (=0.067 M$_{\rm Sun}$) 
    \item $-$0.5, hybrid: 72.3 M$_J$ (=0.069 M$_{\rm Sun}$) 
    \item $-$0.5, cloud-free: 81.7 M$_J$ (=0.078 M$_{\rm Sun}$)    
\end{itemize}

These are slightly lower than the comparable calculations from \citet{Saumon08}, where the cloudy HBMM was 73.4 M$_J$ (=0.070 M$_{\rm Sun}$) and the cloud-free was 78.6 M$_J$ (=0.075 M$_{\rm Sun}$) at solar metallicity. They are also slightly lower than the 78.6 M$_J$ (=0.075 M$_{\rm Sun}$) for solar metallicity, cloud-free evolution found recently by \citet{Chabrier23}. 

\begin{figure*}
    \centering
    \includegraphics[width=3.5in]{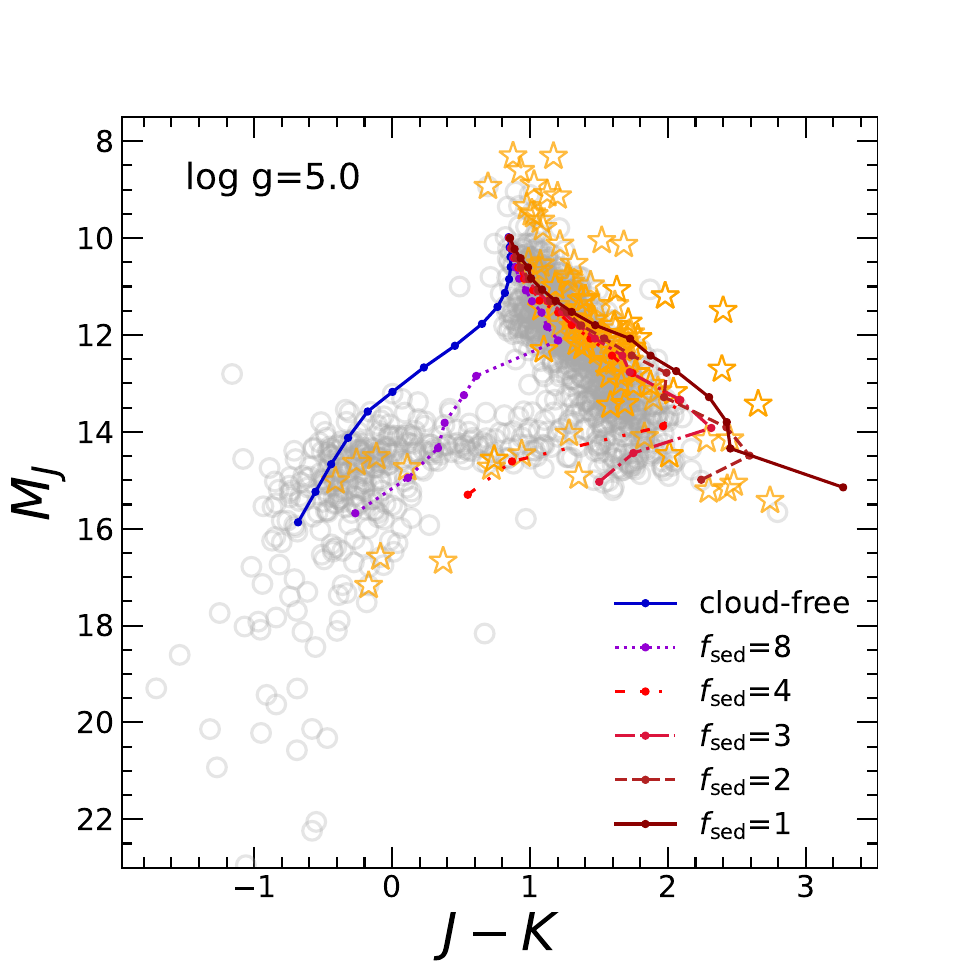}
    \includegraphics[width=3.5in]{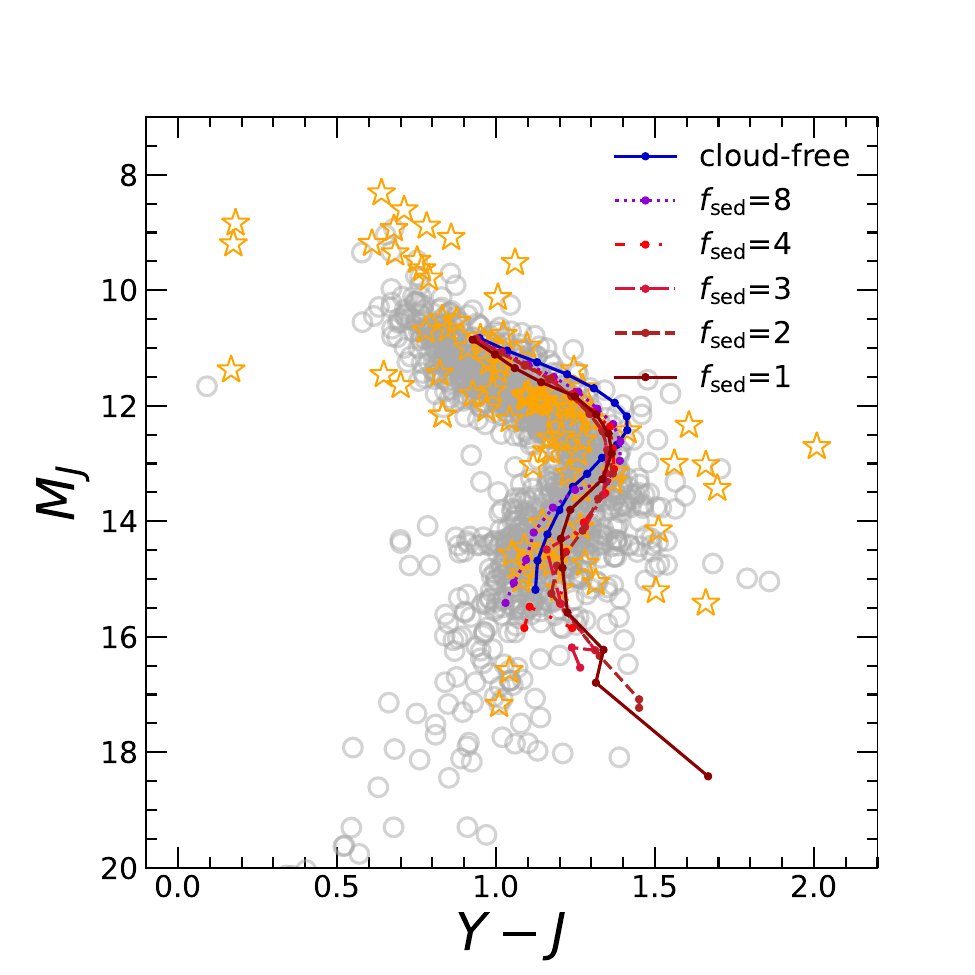}
    \includegraphics[width=3.5in]{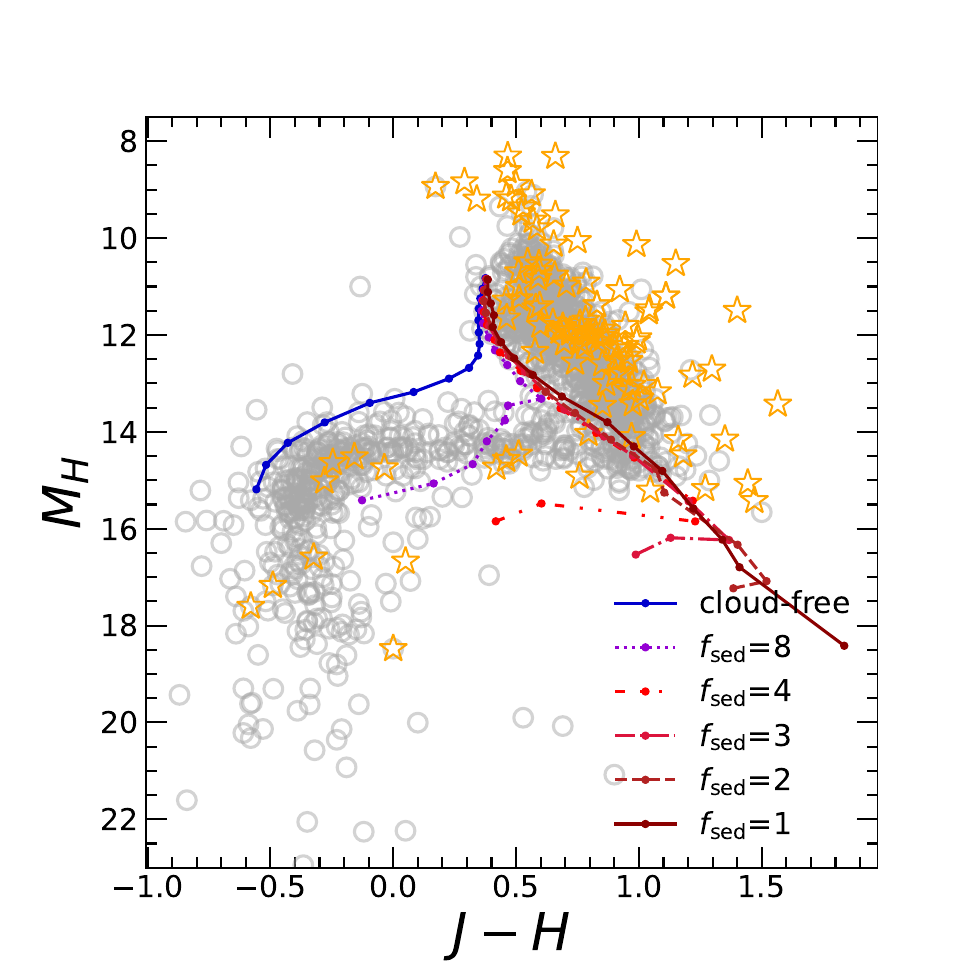}
    \includegraphics[width=3.5in]{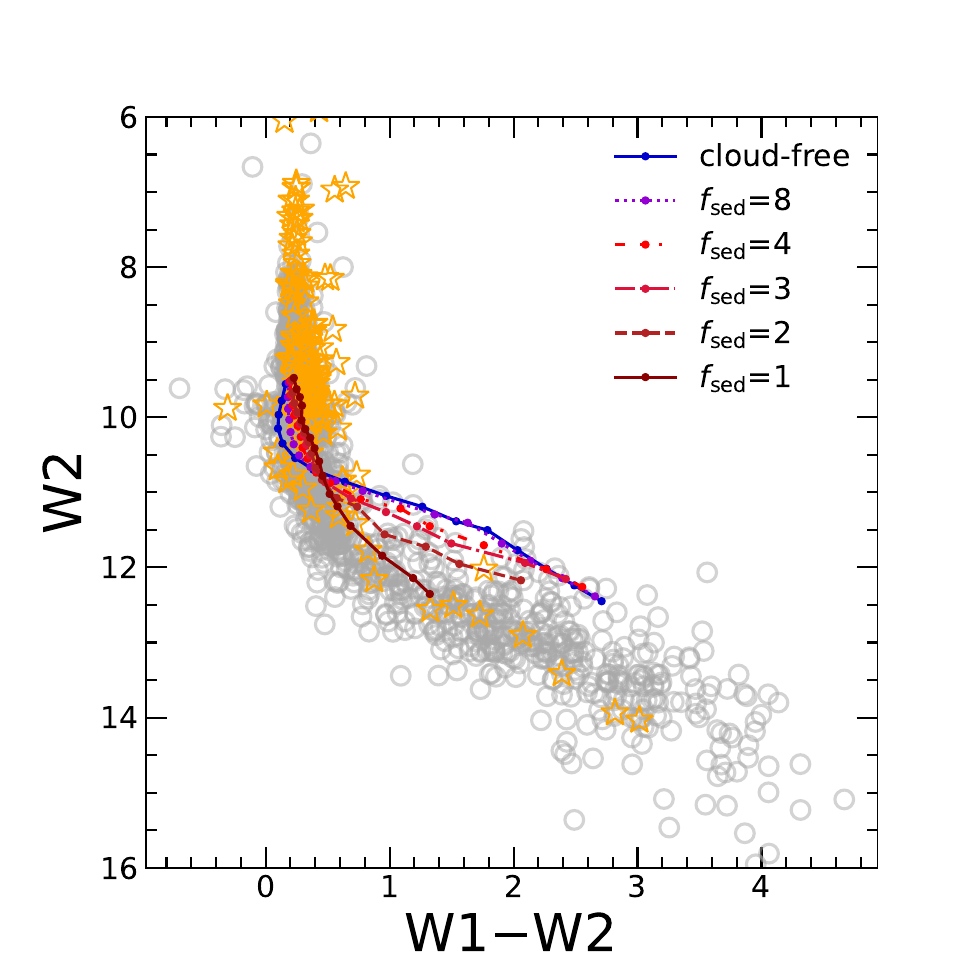}
    \caption{Color-magnitude diagrams illustrating the effect of clouds. Cloud-free models are blue, and cloudy models are purple and red. The gray circles show the photometry of field brown dwarfs; the orange stars show the photometry of `young' brown dwarfs (see text).  All models shown have log g=5.0. No single set of models accurately capture the observed photometry of all brown dwarfs, but in general, cloudy models fit the photometry of L dwarfs more accurately, while cloud-free models fit the photometry of T dwarfs more accurately. }
    \label{fig:cmd_4panel}
\end{figure*}

\begin{figure*}
    \centering
    \includegraphics[width=3.5in]{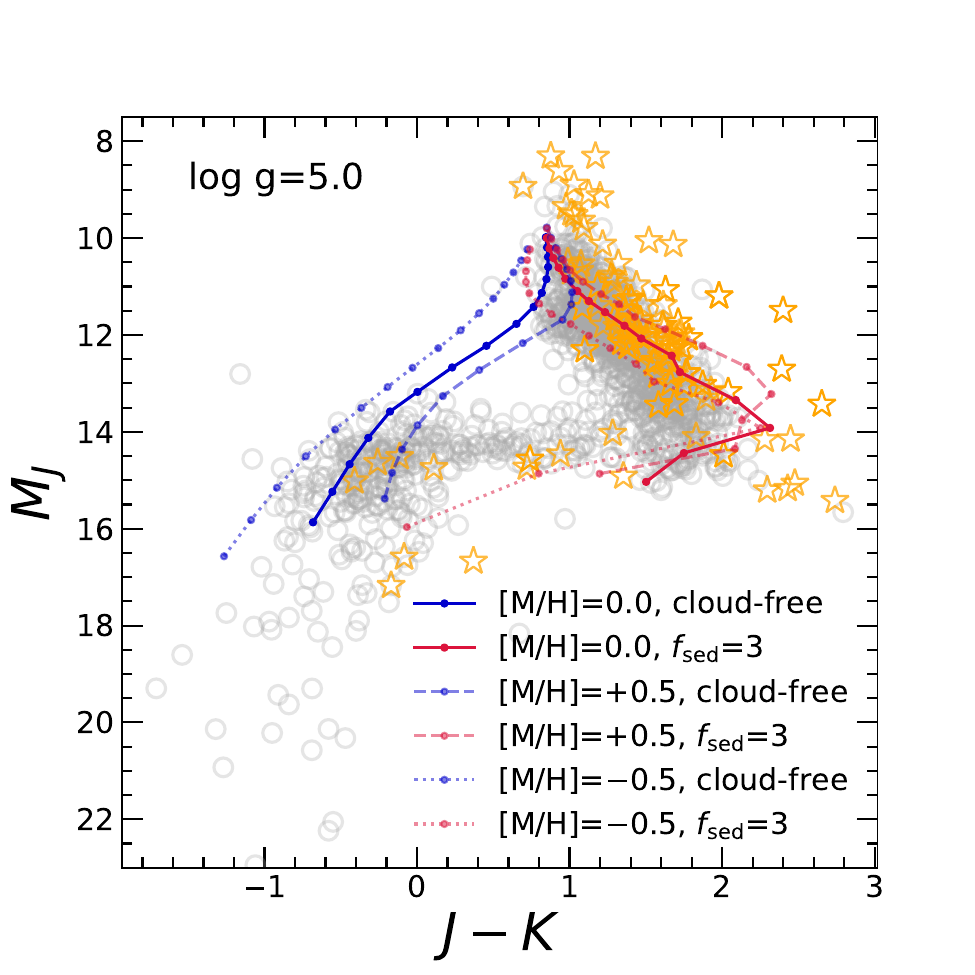}
    \includegraphics[width=3.5in]{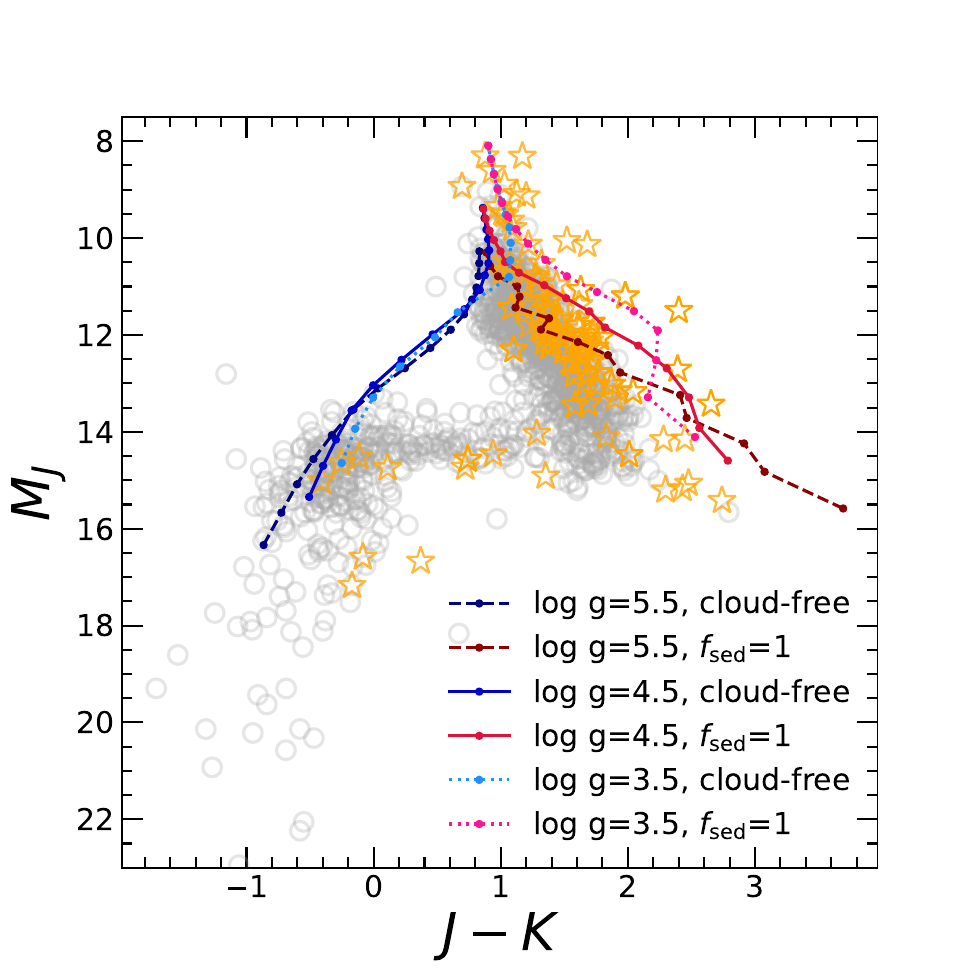}

    \caption{Near-infrared color-magnitude diagram, illustrating the effect of metallicity and surface gravity. In both panels, blue lines show cloud-free models and red lines show cloudy models. The gray circles show the photometry of field brown dwarfs; the orange stars show the photometry of `young' brown dwarfs (see text). \emph{Left:} Solar metallicity models are show as solid lines, while $-0.5$ and $+0.5$ are shown as dashed and dotted lines respectively. For cloud-free atmospheres, lower metallicity models are always bluer in near-infrared colors for a given temperature. For cloudy models, lower metallicity models are typically bluer, with some more complex behavior as the clouds thicken and sink below the near-infrared photosphere.  \emph{Right:} Models are shown for three surface gravities: log g=5.5 as dashed lines, log g=4.5 as solid lines, and log g=3.5 as dotted lines. Lower gravity objects are larger and more luminous (moving up on the CMD) and typically somewhat redder, though the behavior at cooler temperatures shows additional complexity. }
    \label{fig:cmd_met}
\end{figure*}

\subsection{Color-Magnitude Diagrams}

We calculate synthetic photometry for the grid of atmosphere models, using the evolution models to calculate the radius for a given \teff\ and log g. We compare these to the population of brown dwarfs, using the compilation of brown dwarf data presented in the UltracoolSheet \citep{Dupuy12, Dupuy13, Liu16, Best18, Best21}. Field brown dwarfs are shown in gray while any object that is `young' is colored orange. We use both the Banyan-$\Sigma$ Bayesian probabilities of young moving group membership \citep{Gagne18a, Gagne18b} and `flags' in the UltracoolSheet data to assess the youth, including any object with either a Banyan-$\Sigma$ probability greater than 97.5\% or a `young' flag. 

Objects with temperatures of 2400 K have approximately the same synthetic photometry regardless of the \fsed\ value; this is because the clouds are very optically thin. As an object cools, a cloud-free object typically stays roughly the same color until 1900 K, before becoming bluer in its near-infrared colors. In contrast, the cloudy models become progressively redder. Thicker clouds (\fsed=1--2) generally have redder colors, with clouds persisting to lower effective temperatures (900--1000 K), while thinner clouds (\fsed=3--4) tend to become less optically thick at the coolest temperatures in our grid as their more compact clouds sink below the photosphere, becoming reddest at 1200--1300 K. \fsed=8 models are typically lightly cloudy, appearing more similar to cloud-free models. 
 
Different metallicities and surface gravities are shown in Figure \ref{fig:cmd_met}. For cloud-free models, the low metallicity models are always bluer than higher metallicity models. This is typically true for the cloudy models as well, with some complexity as the clouds begin to sink below the photosphere (\teff=1000--1300 K). Lower gravity models tend to be larger and therefore brighter (moving up on the CMD) and have similar or somewhat redder colors at most effective temperatures. There is some complexity, especially for cooler L dwarfs where the cloud base is deeper and the cloud is thicker. The reddest objects remain a challenge for even our cloudiest model spectra to fit.

\section{Discussion}


\subsection{Convergence challenges for cloudy brown dwarfs}

Converging self-consistent 1D radiative--convective models of cloudy, warm brown dwarfs can be somewhat fiddly. As shown in Figures \ref{fig:ptprof} and \ref{fig:convzones}, the cloud opacity tends to create additional detached convective regions, and small perturbations to the P/T structures cause changes in the clouds as convergence is found. We mitigate some of these problems by `smoothing' the cloud as the model converges (taking a weighted average of the prior few iterations of the cloud as it converges), but some issues remain. The astute user of the models may notice that a few models do not vary smoothly as a function of the parameters in the grid; these discontinuities should be considered numerical, not physical.

\subsection{Chemical (dis)equilibrium}

We have presented models that assume chemical equilibrium throughout the atmosphere. This assumption is most accurate for the higher temperatures in our model grid, where CO is the dominant carbon-bearing species and N$_2$ is the dominant N-bearing species. As seen in many prior works \citep{Saumon06, Hubeny07, Miles18, Miles23}, as the atmosphere begins to have more CH$_4$ and NH$_3$, the vigor of mixing relative to chemistry begins to matter. This effect appears to be gravity-dependent, with low mass objects more impacted by chemical disequilibrium. The Sonora Cholla \citep{Karalidi21} and upcoming Sonora Elf Owl models (Mukherjee et al., in prep.) include disequilibrium chemistry but no clouds; we leave the combination of clouds and disequilibrium to future work, but users of these models should be aware of this limitation, especially near the L/T transition. 

\subsection{The silicate feature at 9--10 $\mu$m}

As discussed in the introduction, there is strong evidence for a silicate feature at 9--10 \micron, caused by small silicate particles in the upper atmospheres of mid-L dwarfs \citep{Cushing06, Burningham21, Luna21, Vos23, Suarez22, Suarez23}. For the same reasons noted in \citet{Luna21}, the models presented here do not contain a strong silicate feature. The reason is that within the \citet{AM01} cloud modeling framework, the distribution of particles at different altitudes is not a free parameter; it is calculated self-consistently with their rates of lofting and precipitating in the atmosphere with an advective--diffusive balance. No \fsed\ appears to end up with the right distribution of small cloud particles capable of producing the silicate feature: lower \fsed\ values can have small particles at the right altitudes, but become extremely optically thick deeper in the atmosphere, creating near-infrared spectra that look nothing like real brown dwarf spectra, while higher \fsed\ values better reproduce the near-infrared spectra but lack the silicate feature. We suggest that future models further study this problem, likely incorporating more complexity into the cloud modeling. For example, modeling of the cloud microphysics finds bimodal (or even more complex) particle size distributions \citep{Powell18}. Cloud nucleation at a range of heights in the atmosphere may create higher altitude small particles.

\section{Conclusions}

We have presented atmosphere and evolution models appropriate for comparison with observations of warm brown dwarfs and directly-imaged planets with effective temperatures from 900--2400 K. We include a range of surface gravities from log g=3.5 to 5.5, metallicities from $-0.5$ to $+0.5$, and cloud thickness from thin (\fsed=8) to thick (\fsed=1). 

We present evolution models at three metallicities ($-0.5$, 0.0, $+0.5$) that include clouds which clear at the L/T transition, mimicking the behavior observed in the observed population of brown dwarfs. We find that both metallicity and clouds can strongly impact the thermal evolution of planets and brown dwarfs. 

These Sonora Diamondback cloudy models are the next step in our project to update and expand prior generations of atmosphere and evolution models, following the cloud-free, chemical equilibrium Sonora Bobcat models of \citet{Marley21} and cloud-free, chemical disequilibrium of \citet{Karalidi21}. Future models will further expand the treatments of clouds and disequilibrium chemistry based on our growing understanding of substellar atmospheres.

\begin{acknowledgments}

This work benefited from the 2022 and 2023 Exoplanet Summer Program in the Other Worlds Laboratory (OWL) at the University of California, Santa Cruz, a program funded by the Heising-Simons Foundation. 
C.V.M. acknowledges support from the Alfred P. Sloan Foundation under grant number FG-2021-16592.
C.V.M. acknowledges the National Science Foundation, which supported the work presented here under Grant No. 1910969. C.V.M acknowledges support from the NASA XRP program from grant 80NSSC19K0446. This material is based on work supported by the National Aeronautics and Space Administration under grant No. 80NSSC21K0650 for the NNH20ZDA001N-ADAP:D.2 program. 
R. L. acknowledges support from the NASA ROSES XRP program with the grant 80NSSC22K0953, and from STScI grants JWST-AR-01977.007-A and JWST-AR-02232.008-A. N.E.B. and E.G.N. acknowledge support from NASA ROSES XRP program grant 80NSSC22M0096. 
This work has benefited from The UltracoolSheet at http://bit.ly/UltracoolSheet, maintained by Will Best, Trent Dupuy, Michael Liu, Rob Siverd, and Zhoujian Zhang, and developed from compilations by Dupuy \& Liu (2012), Dupuy \& Kraus (2013), Liu et al. (2016), Best et al. (2018), and Best et al. (2021). This research has made use of NASA's Astrophysics Data System Bibliographic Services. 

\end{acknowledgments}


\bibliography{references}{}
\bibliographystyle{aasjournal}



\end{document}